\documentclass[aps,pre,amsmath,amssymb,reprint,twocolumn]{revtex4-2}
\usepackage{graphicx}
\usepackage{enumerate}
\usepackage{amsmath}
\usepackage[caption=false]{subfig}
\usepackage{hyperref}
\usepackage{float}
\usepackage{xcolor}
\usepackage[a-1b]{pdfx}
\usepackage{soul}
\begin{document}

\title{Understanding jump discontinuity in disordered system}

\author{Anjan Daimari}
\author{Diana Thongjaomayum}
\email{dianat@tezu.ernet.in}
\affiliation{Department of Physics, Tezpur University, Assam 784028, India }

\date{\today}

\begin{abstract}
The response of a complex system to a slow varying external force often displays a jump discontinuity in the order parameter near the critical point. However, this discontinuity is not usually a
single jump but rather breaks into smaller jumps which makes it difficult to locate the critical point on approaching its vicinity based only on simulations, in the absence of exact results.
Our work is a small effort in understanding these breaks in jump through the hysteretic response of a classical Ising spin system to an external field, $h$, in the context of a nonequilibrium zero-temperature random field Ising model on dilute systems. We consider a Bethe lattice with
coordination number, $z = 4$, and dilute a fraction $(1-c)$ of the sites. Therefore the lattice now consists of sites with varying $z = 4, 3, 2, 1$ and possibly few isolated sites $(z=0)$, depending on the concentration $c$. We obtain the exact solution of
the magnetization curve, $m(h)$ vs $h$, for the entire lattice as well as for each sublattice of different $z$ coordinated sites, $m_4(h), m_3(h), m_2(h), m_1(h), m_0(h)$. 
The discontinuity in total magnetization is the result of the superposition of the jumps of different $z$ coordinated sites and
observed at the same value of external field, $h_{crit}$. 
The dominant contribution to the jump comes from those sites with higher concentration and larger $z$. However, the triggering sites responsible for large jumps are mostly $z\ge3$. 
We test this on cubic lattices as well, where exact results are not available. We hope our analysis will help in understanding fluctuations around a jump in numerical simulations
as well as experiments.

\end{abstract}

\maketitle 

\section{Introduction}
\begin{figure}[htb]
    \includegraphics[width=8cm, height=5cm]{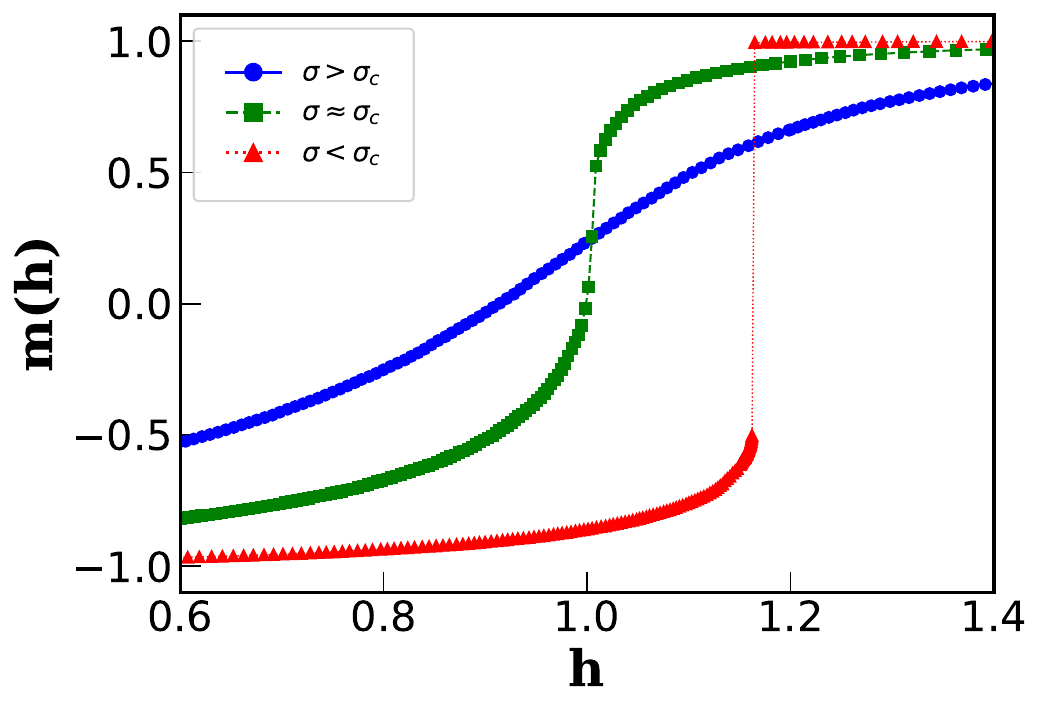}
    \caption{Schematic plot showing smooth magnetization curve in increasing field $h$ for $\sigma > \sigma_c$, large jump for $\sigma < \sigma_c$, and breaks in jumps on approaching $\sigma \approx \sigma_c$ in the context of RFIM.
}
    \label{fig:i1}
\end{figure}

The nonequilibrium zero-temperature random field Ising model (RFIM) has been widely studied to understand disordered magnetic systems\cite{sethna1993hysteresis,wohlman1984random,nattermann1997,sabhapandit2000distribution,sethna2001crackling,vives2000hysteresis,janicevic2017critical,spasojevic2018crossover,mijatovic2021nonequilibrium,bingham2021experimental,rosinberg2009t,illa2006diluted,aharony1985dilute}. In RFIM, each lattice site is assigned a random field, which introduces quenched disorder in the system.
 When such a system is subjected to an external field, the response is governed by transitions between local minima of the rugged free-energy landscape, which are separated by high-energy barriers of various heights. This behavior leads to slow and nonequilibrium relaxation, as the system is trapped in one metastable state before transitioning to other states under the influence of the applied field. 
 A particular point of interest is the jerky response even when the external field is varied smoothly. This response manifests as breaks in the magnetization curve of the system $m(h)$ 
 when the external field $h$ is driven from negative to positive saturation in a cyclic manner, thus resulting in a critical hysteresis loop \cite{dhar1997zero,shukla1996hysteresis,sabhapandit2004absence,bupathy2017random}. 
  There exists a critical disorder $\sigma_c$ below which there are jumps in the $m(h)$ curve that gradually reduce to zero as $\sigma$ increases and result in a smooth $m(h)$ curve when $\sigma>\sigma_c$ \cite{sethna2001crackling,kharwanglang,illa2006diluted,sabhapandit2000distribution}, as shown in Fig. \ref{fig:i1}. 
  The surprising thing is the similarity in the scaling behavior of the avalanches associated with equilibrium and nonequilibrium critical points in the context of RFIM\cite{liu2009unexpected}.
  For the sake of completeness, RFIM was first introduced in a thermal equilibrium system to study the effect of quenched disorder on phase transition\cite{imry1975random}. 
  Although for the equilibrium case, no hysteresis is observed as the system is in global free energy minima unlike their nonequilibrium counterparts where the system is in local minima, with the driven field and shows hysteresis. Nonetheless, the similarity provides a way to understand equilibrium properties through the study of nonequilibrium systems and vice versa. Our work focuses on nonequilibrium systems only as for the equilibrium RFIM other algorithms need to be implemented which is beyond the present study.

   Nonequilibrium RFIM was introduced by Sethna \textit{et al.} to explain Barkhausen noise and return-point memory in driven magnetic systems. 
   They employ the mean-field approach to show the universality of the critical point and power-law behavior in avalanche distribution. Later exact solution of RFIM in one-dimensional (1D) Ising chains was obtained\cite{shukla1996hysteresis,shukla2000exact}. Subsequently, 
   an exact solution on integer $z$ coordinated Bethe lattices shows $\sigma_c$ exists only when $z\ge3$\cite{dhar1997zero}. These findings raise the question of dependence of $z$ in addition to dimensionality $d\ge2$ for the observance of jump discontinuity. In the equilibrium case, the lower critical dimension is found to be $d=2$\cite{imbrie1984lower}. Recent studies also show the crucial presence of $z=4$ sites in mixed $z=3$ and $z=4$ Bethe lattices
for the occurence of jump discontinuity\cite{diana2016hysteresis}.

The role of $z$ in RFIM has also been explored by studying 
diluted systems viz. triangular lattices and Bethe lattices.
   In addition to randomness, dilution is also a key factor in ensuring the presence of sufficient participating sites in the avalanche when other criteria for jump discontinuity are fulfilled \cite{kharwanglang,diana2017criteria}. A regular $z=3$ honeycomb lattice does not show critical behavior\cite{sabhapandit2002hysteresis}, whereas the $z=4$ square 
 lattice\cite{spasojevic2011numerical} and the $z=6$\cite{diana2013effect,diana2019critical} triangular lattice have a critical response. A gradual dilution scheme with $c$ as the concentration of sites present in a $z=6$ triangular lattice can result in $z=3$ honeycomb when $c\rightarrow0$. Numerical studies show when $c<1/3$, which corresponds to effective coordination number, $z_{eff}<4$, the signatures of spanning avalanches are absent in the diluted triangular system \cite{kurbah2015nonequilibrium}. 
   Further, studies on diluted $z=4$ Bethe lattice with concentration $c$ of the present sites show that, when $c>0.557$ and for any $\sigma<\sigma_c(c)$, critical hysteresis has been observed\cite{diana2017criteria}.
Despite several earlier works, the main challenge that remains in systems exhibiting jump discontinuity is to locate the critical point specified by critical disorder $\sigma_c$ at $h_{crit}$, where the jump just disappears. In the case of Bethe lattices, where analytical results are available, it could be determined easily. It would be in order to mention that the analytical results obtained on Bethe lattices for the undiluted or diluted case could be verified through simulations on random graphs. Even in this case, data from simulations alone always shows jumps with several breaks on approaching $\sigma_c$, which does not disappear even on increasing the system size\cite{dhar1997zero,kharwanglang,diana2016hysteresis}. Therefore, determining $\sigma_c$ only through simulations or experiments remains a daunting task. 

 We plan to delve into the make-up of the jump discontinuity in terms of contributing different $z$ sites on dilute $z=4$ Bethe lattices in the framework of RFIM. Our approach is to understand the source of the fluctuations and the fragmented jumps through the analysis of magnetization from different $z$ coordinated sublattices. This is a new perspective and definitely extends the existing works in the field discussed earlier. A diluted $z=4$ Bethe lattice consists of sites with different $z=4,3,2,1$ and even $z=0$ isolated sites. We analyze the overall magnetization behavior of the entire lattice as well for each sublattice with coordination numbers of $4, 3, 2, 1$, and $0$.  We obtain exact solutions of the magnetization curves for the entire lattice and individual sublattices with different $z$. 
The aim is to assess whether the observed jumps in magnetization curve are exclusively associated with coordination numbers 4 or if they extend to other lattice sites and whether this is observed at the same or different values of $h$. 
Our studies show that the fluctuations in the jump on approaching the critical point are collectively driven by several jumps of different $z$ sites superimposed on one another and occurs at the same value of $h=h_{crit}$.
We perform numerical simulations on cubic lattices for $\sigma<\sigma_c\approx 2.272$ \cite{fytas2013universality} and examine the correspondence of the results obtained from Bethe lattices. We find similar trends governing the jump discontinuity in the cubic case where the analytical result is not plausible.

\section{Model}
The dilute random-field Ising model (RFIM) is defined by the Hamiltonian
\begin{equation}
 H = -J \sum_{\langle i,j \rangle} c_i c_j S_i S_j - \sum_i h_i c_i S_i - h \sum_i c_i S_i  
\end{equation}
where \( J > 0 \) is a ferromagnetic exchange interaction and the sum runs over nearest-neighbor sites \(\langle i,j \rangle\) of a lattice. 
Here, \( S_i = \pm 1 \) represents an Ising spin and \( h_i \) is a random field at site \(i\); \( h \) is a uniform external field. The random field \( h_i \) is drawn from a Gaussian distribution with zero mean and standard deviation \( \sigma \). The quantity \( c_i \) is a random variable at site $i$ that takes the value 1 with probability \( c \) of being occupied, or 0 otherwise. Thus \( c \) represents the concentration of occupied lattice sites. The variables \( \{ h_i \} \) and \( \{ c_i \} \) are quenched, remaining fixed throughout the system's evolution, while the spins \( S_i \) are the dynamical variables. These spins follow the zero-temperature Glauber dynamics at discrete time steps \( t \)\cite{glauber1963time}:
\[
S_i(t+1) = \text{sgn} \left[ J \sum_j c_j S_j(t) + h_i + h \right]\tag{2}
\]
The sum on the right-hand side runs over nearest neighbors of site \( i \). For a fixed applied field \( h \), the system iteratively lowers its energy and converges to a stable fixed point \( \{ S_i^*(h) \} \), such that for each lattice site \( i \)
\[
S_i^*(h) = \text{sgn} \left[ J \sum_j c_j S_j^*(h) + h_i + h \right] \tag{3}
\]
A stable fixed point is reached when all the spins are aligned along their respective local fields for the fixed value of $h$. Usually we start with an adequate negative \(h\) such that all spins \( S_i(h = -\infty) = -1 \). The external field is then slowly increased until a spin becomes unstable. In this stage, the field is held fixed, and the system is allowed to evolve under the iterative dynamics until it reaches a new fixed point. This new fixed point is characterized by the magnetization per site \( m(h) \) for total $N$ sites: 
\[
m(h) = \frac{1}{cN} \sum_i c_i S_i^*(h), \tag{4}
\] 
On sweeping the entire system from $h=-\infty$ to $+\infty$ with minimal increase in $h$ so as to flip the next unstable spin, we get the lower hysteresis loop. The magnetization for each $z$ coordinated sublattice can be calculated separately using
\[m_z(h)=\frac{1}{cN}\sum_i c_iS_i^*(h)\]
where the summation is only over those sites with the coordination number $z$. We shall focus only on the lower hysteresis loop as the upper hysteresis loop obtained on varying $h=+\infty$ to $-\infty$ is symmetric.

\section{Analytical Solution on Bethe Lattice}

\begin{figure}[htb]
   \centering
    \includegraphics[width=6cm, height=3cm]{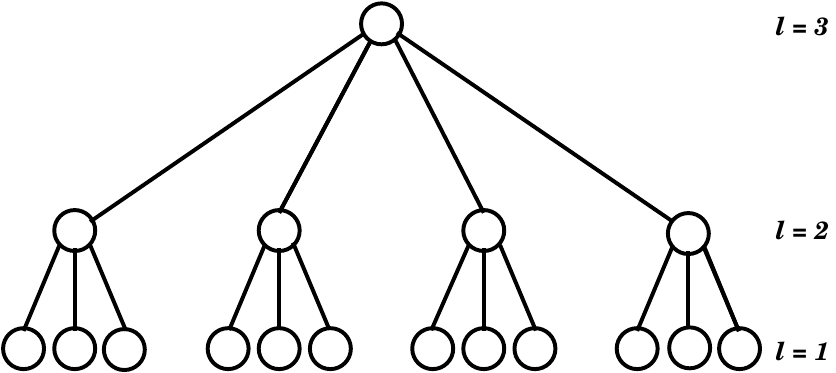}
    \caption{Bethe lattice with coordination number \(z=4\).
}
   \label{fig:i2}
\end{figure}

A Bethe lattice is defined as the innermost part of a Cayley tree of the same coordination number, $z$. Figure~\ref{fig:i2} illustrates a three-level $z=4$ Cayley tree where the lattice points on the bottom row $(l=1)$ form the edge of the tree, and the point at the peak $(l=3)$ represents the core of the tree. As we can see from the figure, except the points on the edge that have $z=1$, all other points have $z=4$. There is no closed loop in the lattice, making it possible to derive analytical results on the Cayley tree using the probabilistic method of spins relaxation at each level $l$. In the limit of large $l$, the surface effects become negligible and the result validates to the deep interior region. Therefore, we employ recursion relations starting from the surface and tend to move towards the interior of the tree. However, simulations are done on random graphs for the same $z$ as it is similar to the interior of the Cayley tree and also helps in getting away with the large lattice points  at the surface. Simulations on the random graph are discussed in the next section.

For the analytical calculation, we generalize the conditional probabilistic approach in Ref.\cite{kharwanglang} to determine the magnetization on a dilute Cayley tree, where each site can be occupied by an Ising spin or left vacant depending on the value of $c$. As the recursive approach works effectively in the dilute Cayley tree, we start with adequately large negative external field $h$ such that the initial configuration of spins are oriented downward. Then, the external field is incremented to a value $h$. Thereafter, spins are relaxed from the surface towards the interior.
Our focus is on the central site that is randomly selected, and calculate the probability of the spin at the central site being occupied and up at $h$. 

As in Ref.\cite{kharwanglang}, we define $P_l(h)$ as the conditional probability of a spin randomly chosen at level $l$ being up, keeping the spin at level $(l+1)$ (parent spin) down and while all the spins at level $(l-1)$ (descendant spin) are fully relaxed. The recursion relation is

\begin{align*}
  P_l(h) =&\sum_{i=0}^{z-1} \binom{z-1}{i} (1-c)^i\Bigg[\sum_{j=0}^{z-i-1} \binom{z-1-i}{j} (cP^{l-1}(h)) ^j\\& \times (c - cP^{l-1}(h))^{z-1-i-j} p_j(i; h)\Bigg], \label{cp} \tag{5}
\end{align*}
where
\begin{align*}
p_j(i; h) = \int_{\left(z-2j-i\right)J-h}^{\infty} \frac{1}{\sqrt{2\pi \sigma^2}} \exp\left(-\frac{h_i^2}{2\sigma^2}\right) \, \mathrm{d}h_i.  \tag{6}
\end{align*}
$p_j(i; h)$ represents the probability that a site at level $l$ has enough random field to flip up at an external field $h$, when $j$ neighbors are up, $i$ neighbors are vacant, and, therefore, $z-j-1$ neighbors are down. $\sigma$ is the standard deviation of Gaussian distribution from where the random field $h_i$ is drawn. 
\\
As we continue to iterate Eq.(\ref{cp}), i.e., when $l\rightarrow\infty$, $P_l(h)$ converges to a fixed point $P^*(h)$, which satisfies the self-consistent equation: 
\begin{align*}
  P^*(h) =&\sum_{i=0}^{z-1} \binom{z-1}{i} (1-c)^i\Bigg[\sum_{j=0}^{z-i-1} \binom{z-1-i}{j} (cP^*(h)) ^j\\& \times (c - cP^*(h))^{z-1-i-j} p_j(i; h)\Bigg]. \label{cpstar} \tag{7}
\end{align*}

The probability of the central site being occupied and up at $h$ is
\begin{align*}
p(h) =&c \sum_{i=0}^{z} \binom{z}{i} (1-c)^i \Bigg[ \sum_{j=0}^{z-i} \binom{z-i}{j} ( cP^*(h))^j \\& \times ( c - cP^*(h))^{z-i-j} p_j(i, h)\Bigg]. \label{cph} \tag{8}
\end{align*}
The magnetization for the entire lattice with concentration $c$ is given by:
\begin{align*}
  m(h)=&2p(h)-c
 \tag{9}
\end{align*}
Consequently, the magnetization of sublattices of $z = 4, 3, 2, 1$ and $0$ coordinated sites can be rewritten from Eq.(\ref{cph}) by substituting $z$ for each sublattice:

\begin{align*}
p_z(h) =&c \binom{4}{4-z} (1-c)^{4-z} [ \sum_{j=0}^{z} \binom{z}{j} ( cP^*(h))^j \\& \times ( c - cP^*(h))^{z-j} p_j(4-z, h)]  \tag{10}
\end{align*}
Magnetization for each $z$ coordinated sublattice is
\begin{align*}
  m_z(h)=&2p_z(h)-c(z)
 \tag{11}
\end{align*}
The concentration of different coordinated sites $c(z)$ is taken from the simulation while comparing with the theoretical results.

\section{Results}
\subsection{Simulation on random graph}
We start with an undiluted $z=4$ random graph with total sites $N=10^6$ and then gradually dilute the sites with probability $(1-c)$. Consequently, the diluted random graph $z=4$ now consists of sites with varying coordination numbers $z=4,3,2,1$ and also a few isolated sites ($z=0)$. 
\begin{figure}[t]
    \centering
    \includegraphics[width=8cm, height=5cm]{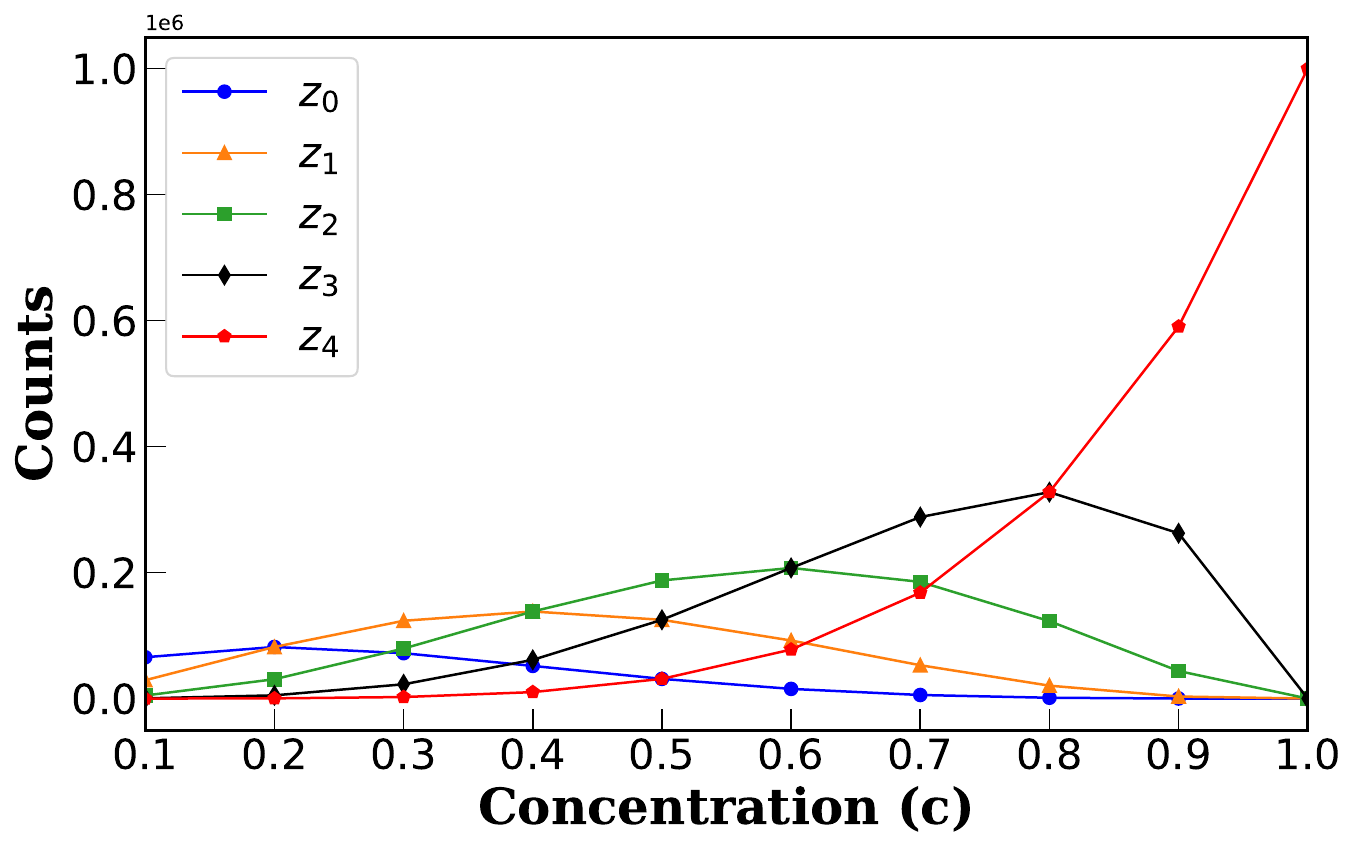}
    \caption{Plot shows the concentrations of different $z$ coordinated sites on varying the total lattice concentration $c$ from 0.1 to 1, present on \(z=4\) random graph with total sites $N=10^6$ averaged over 10 000 data set.
}
    \label{fig:i3}
\end{figure}
Figure ~\ref{fig:i3} shows the concentration of different $z$ coordinated sites on changing the concentration of the entire lattice, $c$ from 0.1 to 1 averaged over a $10^4$ data set. We observe that, as the dilution increases (or $c$ decreases), the total number of \( z_4 \) sites decreases while the concentrations of \( z_3 \), \( z_2 \), \( z_1 \), and \( z_0 \) sites increase and reach their peak values and then decrease ultimately when $c$ becomes less.
A gradual shift in peak concentrations to the left for decreasing $z$ is observed as shown in Fig.~\ref{fig:i3}. We could relate the fact that the propagation of avalanche requires a connectivity of sites from the edge to the central site on the Cayley tree and therefore a minimum concentration $c=1/3$ is required for percolating clusters to survive. 

\begin{figure}[hbt]
    \centering
    \includegraphics[width=8cm]{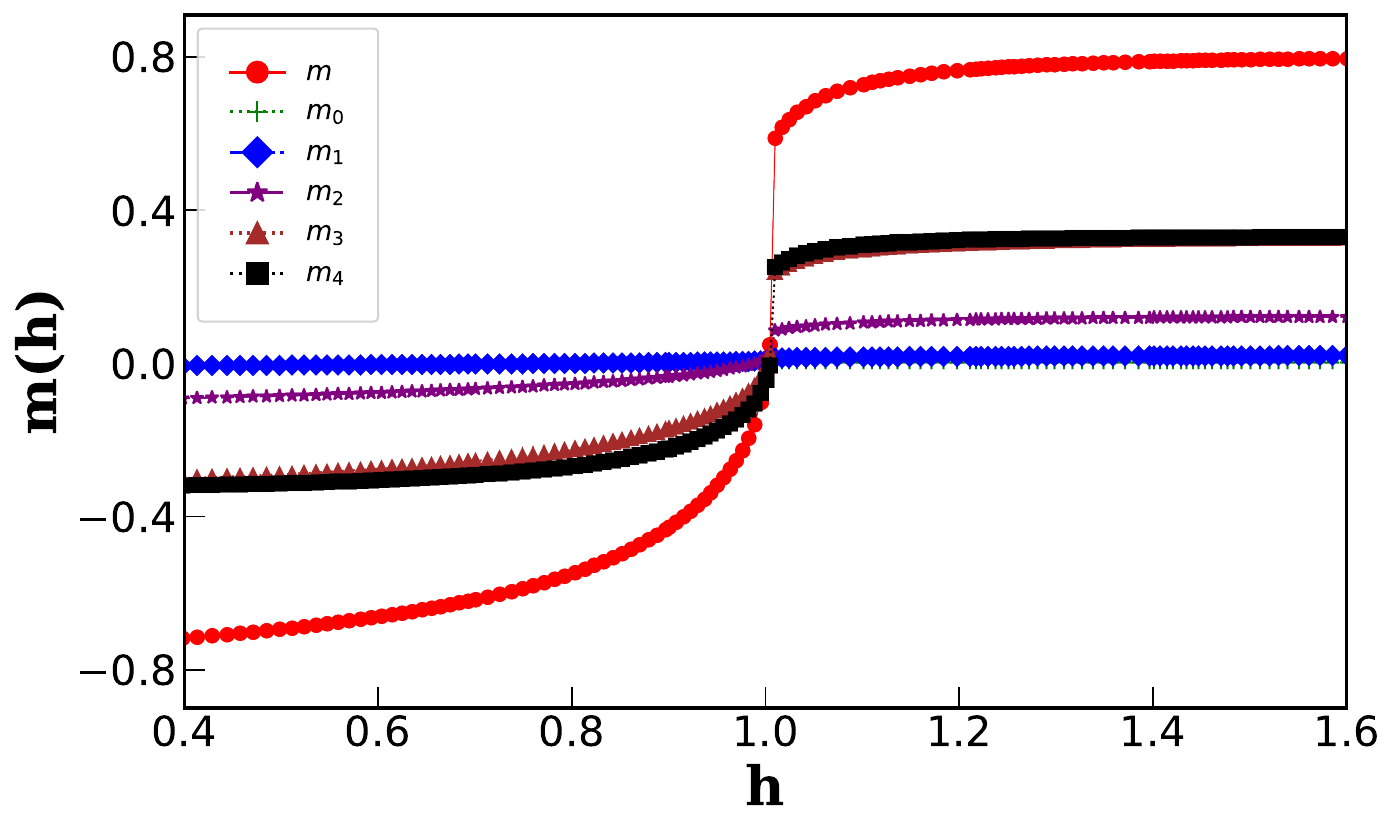}
    \caption{Plot of magnetization, $m(h)$ vs. external field $h$ on random graph with total nodes $N= 10^6$, $c = 0.8$, and $\sigma = 1.2$. The total jump discontinuity of the entire lattice (largest jump denoted by red) is obtained by superposition of jump discontinuities from sublattices with $z = 4,3,2,1$ and occurs at the same value of $h$.
}
    \label{fig:i4}
\end{figure}

We know from earlier studies that jump discontinuity is observed on dilute $z=4$ Bethe lattices when $c>0.557$ for small $\sigma<\sigma_c$. Reference \cite{diana2017criteria} further asserts the crucial presence of few $z=4$ sites, even though other $z<4$ coordinated sites are present, for the propagation of an infinite avalanche. From Fig. \ref{fig:i3}, we notice $z_4$ concentration drops rapidly around $c=0.5$ and subsequently for other $z$ coordinated sites. However, the number of isolated sites $z_0$ tends to grow slowly on crossing $c=0.6$ onward. Thus, at $c_{crit}=0.557$, although the present study 
shows a presence of few $z_4$ sites, they may be too few to sustain the jump in magnetization with  no contribution from the isolated sites and dangling bonds from $z_1$ sites.

\begin{figure}[htb]
    \centering
    \includegraphics[width=8cm]{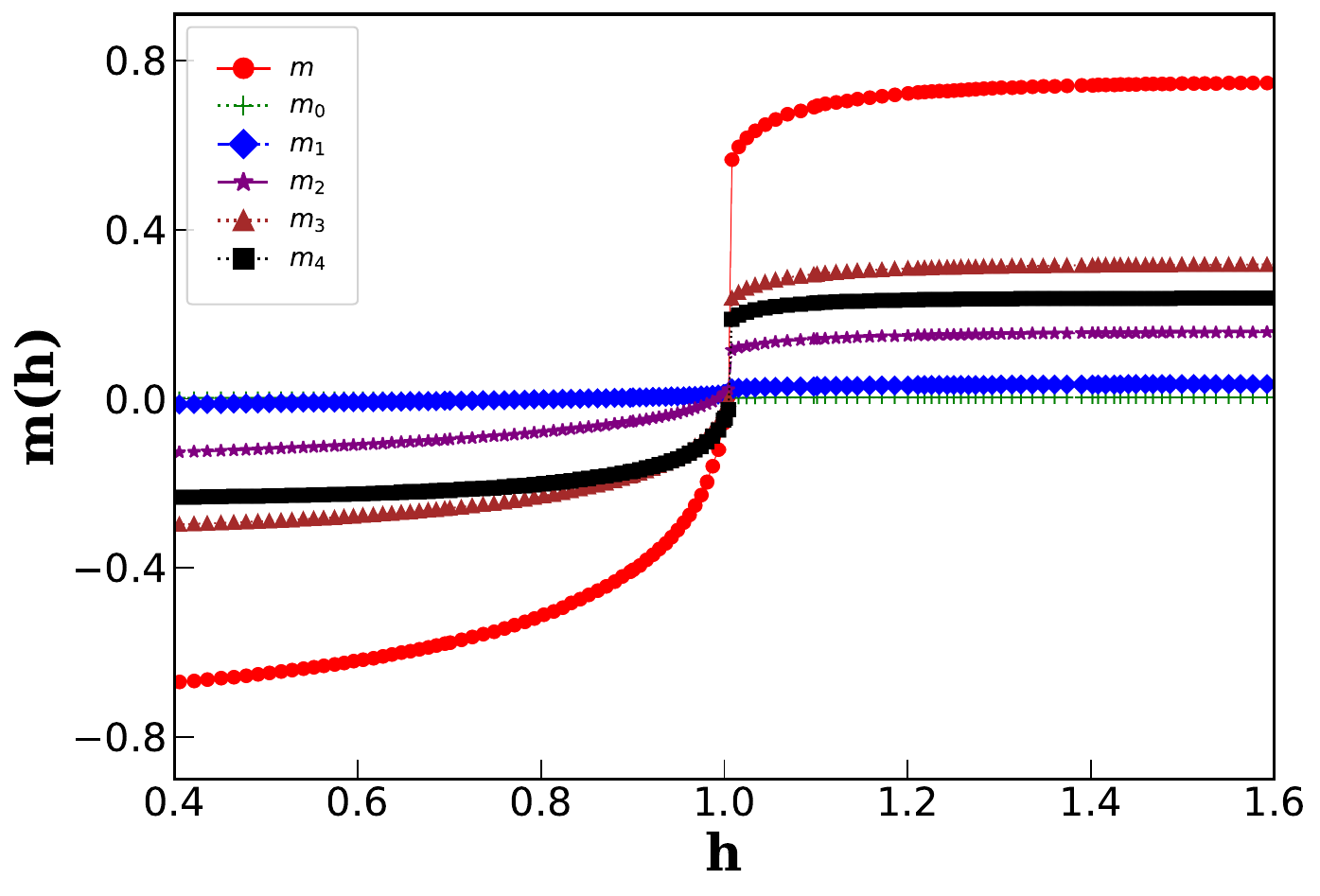}
    \caption{At $c=0.75$, $z=3$ sites have the highest concentration as shown in Fig.~\ref{fig:i3}. This plot confirms that the $z_3$ sublattice has the largest contribution  to the magnetization.
}
    \label{fig:i6}
\end{figure}

In order to understand the contribution of sites with different $z$ to the jump in magnetization, we calculate separately the magnetization in increasing external field $h$ due  to \( z_4 \), \( z_3 \), \( z_2 \), \( z_1 \), and \( z_0 \) sites denoted as $m_4(h)$, $m_3(h)$, $m_2(h)$, $m_1(h)$, and $m_o(h)$, respectively.
To demonstrate, we choose $c=0.8$ and $\sigma=1.2<\sigma_c(=1.6)$ on random graphs with total sites $N=10^6$. 
The total magnetization curve denoted by $m$ and the contributing magnetization from each $z$ coordinated lattice site denoted by $m_0, m_1, m_2, m_3, m_4$, respectively, for $z=0, 1, 2, 3, 4$ are plotted in Fig~\ref{fig:i4}. We purposely choose the value of $\sigma=1.2$ where discontinuity in $m(h)$ curves show up. The analysis shows that the jump discontinuity is not exclusively due to \( z=4 \)sites, but rather \( z=3, 2, 1\) sites also contribute to the jump discontinuity as shown in Fig.~\ref{fig:i4}. 
\begin{figure}[htb]
    \centering
    \includegraphics[width=8.5cm, height=9cm]{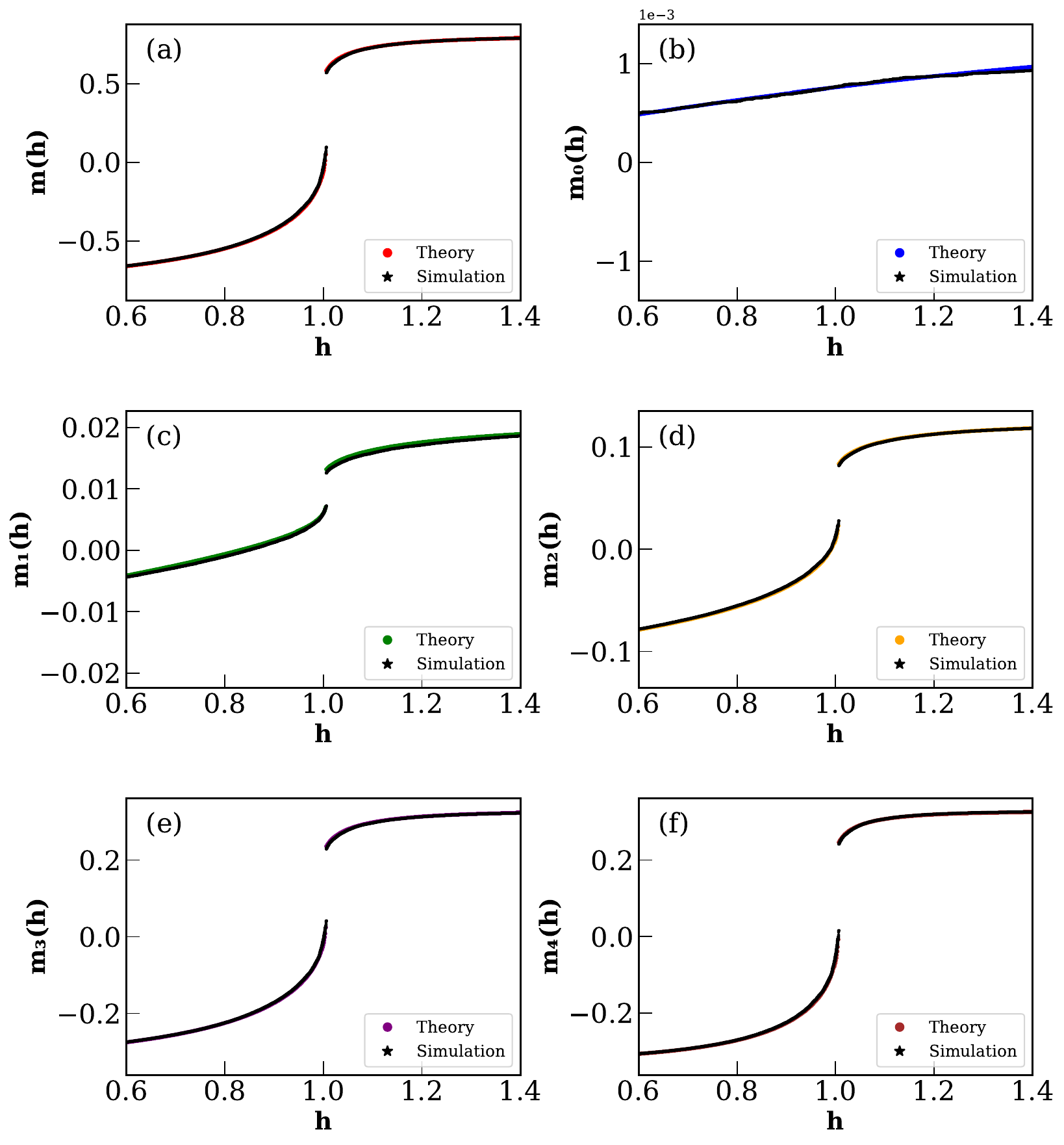}
    \caption{Plots of $m(h)$ vs. $h$ obtained from simulations on random graph matching exactly with the theoretical predictions: (a) corresponds to the whole lattice, and (b)-(f) corresponds to $z=0,1,2,3,$ and $4$ sublattices, respectively. The simulation data is obtained for system size $N= 10^6$, $c = 0.8$, and $\sigma = 1.2$.   
}
    \label{fig:i5}
\end{figure}
Our simulation results exactly match with the theoretical predictions on the Bethe lattice as shown in Fig~\ref{fig:i5} for $c=0.8$, $\sigma=1.2$, and $N=10^6$. The subfigure(a) shows $m(h)$ vs $h$ for the entire lattice where the theory and simulations agree, followed by magnetization curves (theory and simulation) for sublattices of coordination numbers
(b)$z_0$, (c)$z_1$, (d)$z_2$ (e)$z_3$, and (f)$z_4$. For $c=0.8$, $z_4$ concentration is the highest and hence largest contribution to the jump in magnetization curve from a sublattice consisting of $z_4$ sites. We further check the nature of the contribution for $c=0.75$, where $z_3$ sites are more compared to $z_4$ sites. As expected, we find the contribution is highest from the $z_3$ lattice to the total $m(h)$ curve as shown in Fig.~\ref{fig:i6}.

\begin{figure}[h]
    \centering
    \includegraphics[width=8.5cm, height=9cm]{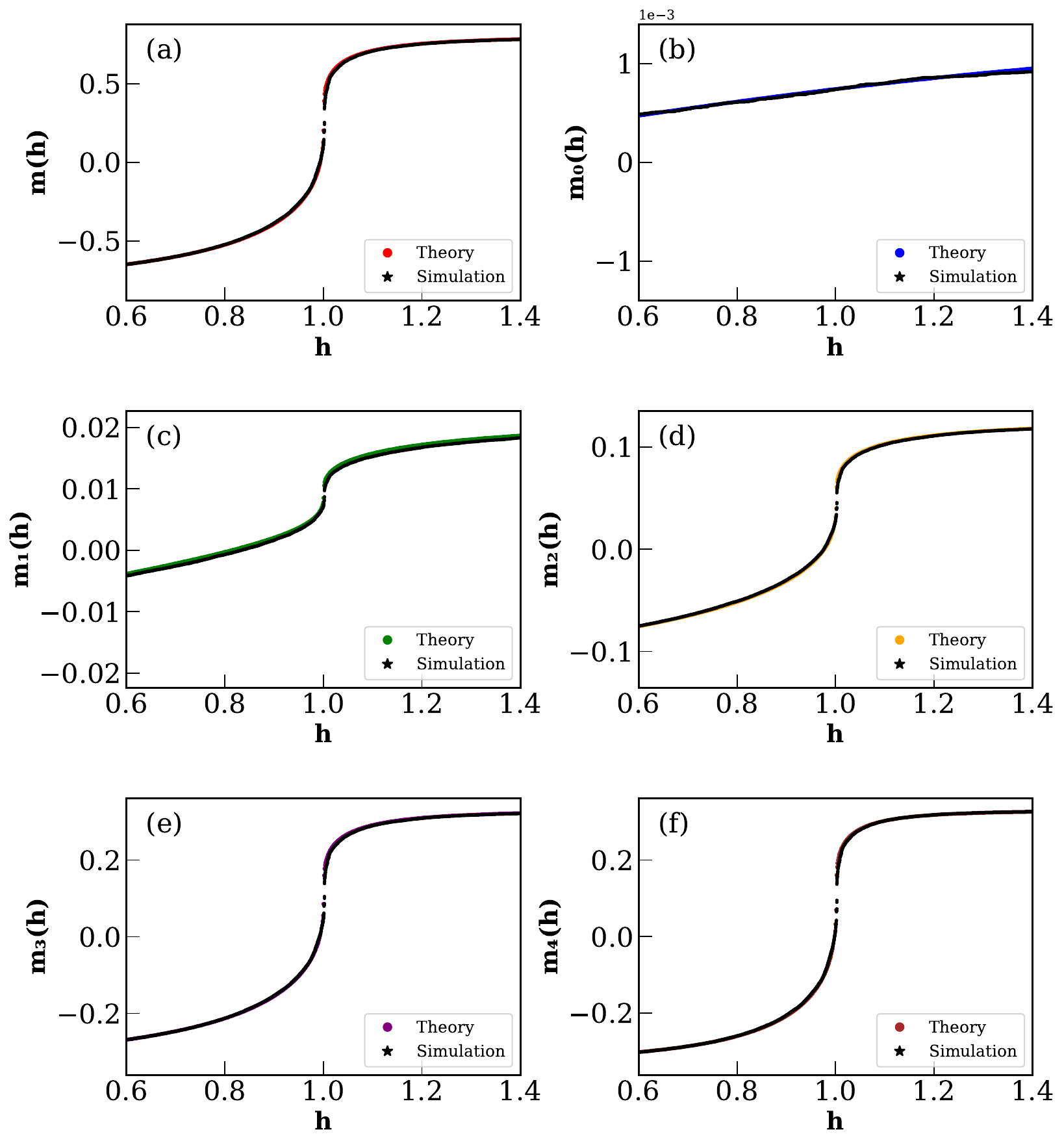}
    \caption{Plots of $m(h)$ vs $h$ at $\sigma \approx \sigma_c$ for both simulation and theoretical predictions, showing several breaks in the magnetization curve (except for $z=0$ sites). Results are shown for the entire lattice in (a), and for $z=0,1,2,3,$ and $4$ coordinated sublattices in (b)-(f), respectively, for system size $= 10^6$ and $c = 0.8$. 
}
    \label{fig:i7}
\end{figure}

Figure ~\ref{fig:i7} shows $m(h)$ vs $h$ simulation results on a $N=10^6$ random graph for $c=0.8$ superimposed on the theoretical predictions when $\sigma\approx \sigma_c$ for the whole lattice [in subfigure (a)] and for $z=0,1,2,3,4$ sublattices separately in (b)-(f). Close to the critical point, we see the breaks in the jumps which were shown by the separate sublattices except for $z=0$.

\begin{figure}[hbt]
    \centering
    \includegraphics[width=8cm, height=5cm]{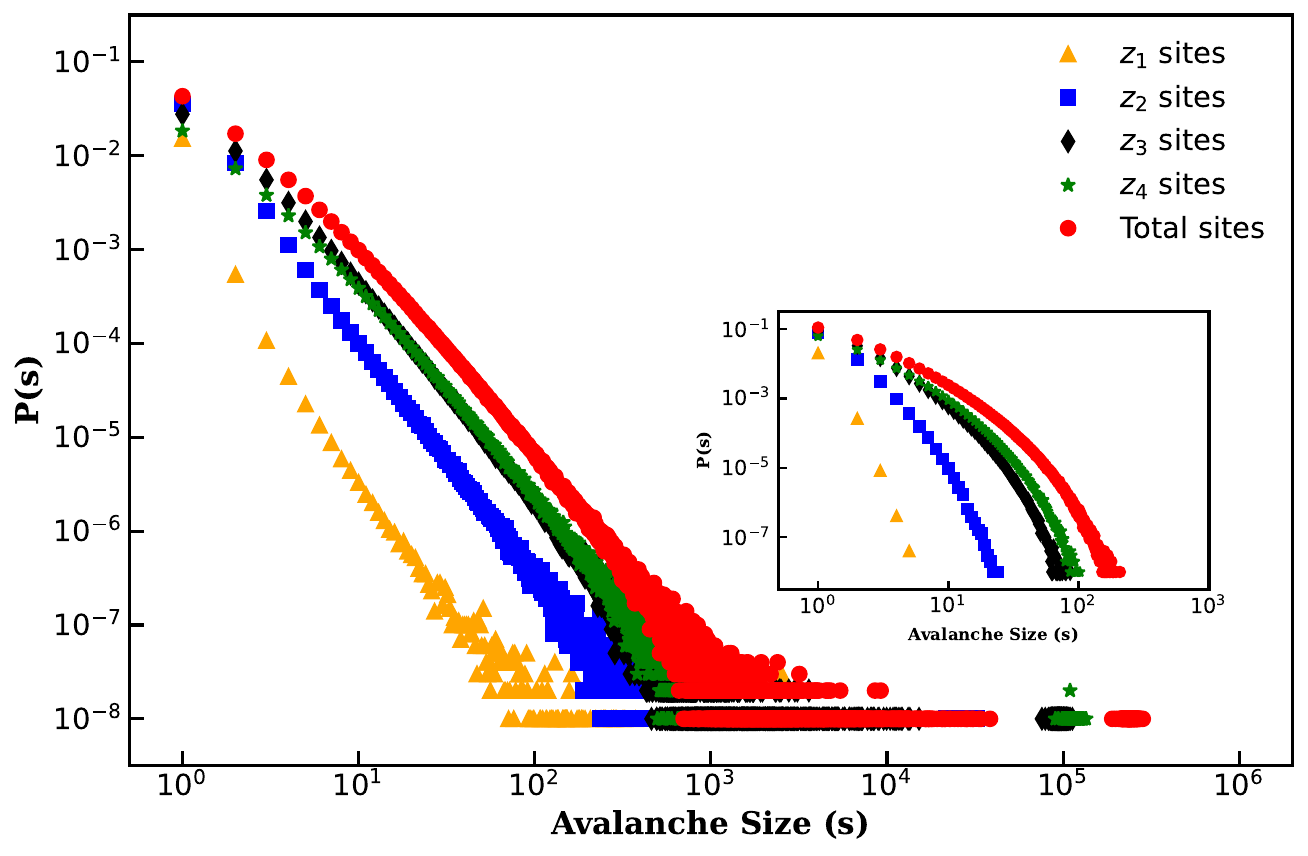}
    \caption{Log-log plots of integrated avalanche size distribution for the whole lattice and sublattices with $z=1,2,3,$ and 4. The data is obtained on a random graph with total sites $N= 10^6$ and $c = 0.8$ averaged over 100 independent configurations. The main figure shows the results for $\sigma=1.2$ and the inset corresponds to $\sigma=2$.  
}
    \label{fig:i7a}
\end{figure}

 A plot of avalanche size distributions in log scale for the whole lattice and different sublattices with $z=1,2,3,$ and 4 on a $N=10^6$ random graph for $c=0.8$ averaged over 100 random field configurations are shown in Figure~\ref{fig:i7a}. We record all the avalanches when the external field $h$ is varied from $-\infty$  to $+\infty$. The data in the main plot is for $\sigma=1.2$ and as expected power law  behavior is observed. The inset also has the same parameters as in the main plot except for different $\sigma=2>\sigma_c$ and hence deviates from the power law, bends down and spans only for few decades of avalanche size.   The power-law exponent $\tau$ [$P(s)\approx s^{-\tau}$] for $\sigma=1.2$ for the whole lattice and $z=1,2,3,$ and 4 sublattices are $2.40\pm0.01$, $1.65\pm0.06$, $2.00\pm0.03$, $2.36\pm0.02$, and $2.32\pm0.02$, respectively. The obtained value of $\tau$ for the entire lattice is consistent with the previous reported results \cite{sabhapandit2000distribution}. 

\begin{figure}[htb]
    \centering
    \includegraphics[width=8.5cm, height=3.8cm]{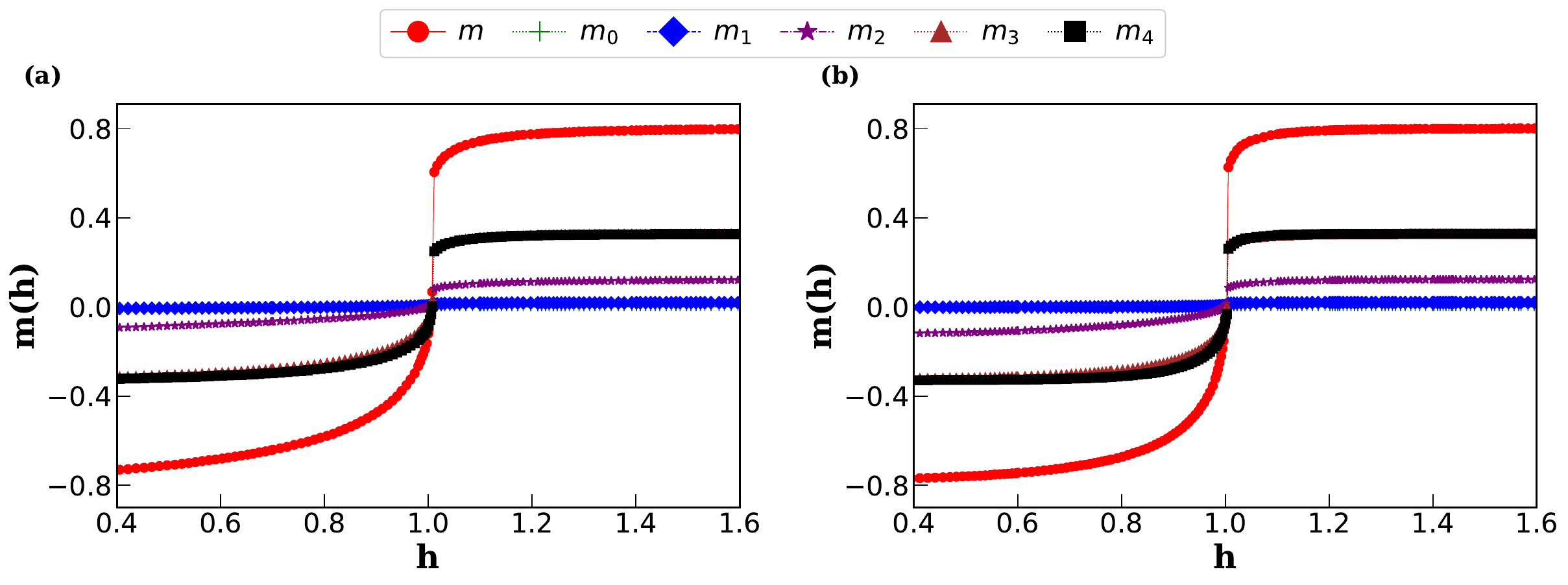}
    \caption{Magnetization $m(h)$ vs external field $h$ on random graph with nodes $N=10^6$ and $c=0.8$ for random field ($h_i$) drawn from (a) uniform distribution, $h_i\in[-2,2]$, and (b) double Gaussian distribution with $\mu_1=-1$, $\mu_2=1$, $\sigma=0.3$, and $p=0.5$.
    }
    \label{fig:i7b}
\end{figure}

To explore whether our results depends on the distribution of random fields, we additionally performed numerical simulations on a random graph ($N=10^6$ and $c=0.8$), with random fields drawn from uniform and double Gaussian distributions. We observe a similar jump in the magnetization curve for both the distributions, which is consistent with our findings. The results are shown in Figure~\ref{fig:i7b} for (a) uniform distribution [-2,2] and (b) double Gaussian distribution with $\mu_1=-1$, $\mu_2=1$, $\sigma=0.3$ and $p=0.5$. The choice of the mixing proportion $p=0.5$ is to ensure symmetry in the distribution; however, the magnetization behavior remains qualitatively the same for both the symmetric and asymmetric ($p\ne0.5$) case (although not shown here). Earlier studies also suggest that symmetry in the distribution of the random field does not play a significant role in hysteresis phenomena \cite{sabhapandit2002hysteresis}.

\par

We also note that the random fields drawn from discrete bimodal ($\pm h_i$) and binomial distributions would contribute to discrete local fields which results in macroscopic avalanches (many spins will flip at the same external field $h$). These large jumps lead to very few $h$ values and hence we find rectangular $m(h)$, curve although the superposition of jumps from sublattices still work. A detailed study on these is beyond the present study and left as future work.
For further discussions on the avalanche distribution on discrete as well as continuous distribution of random fields refer to \cite{sabhapandit2000distribution}.

\subsection{Simulation on cubic lattice}

\begin{figure}[h]
    \centering
    \includegraphics[width=8cm, height=5cm]{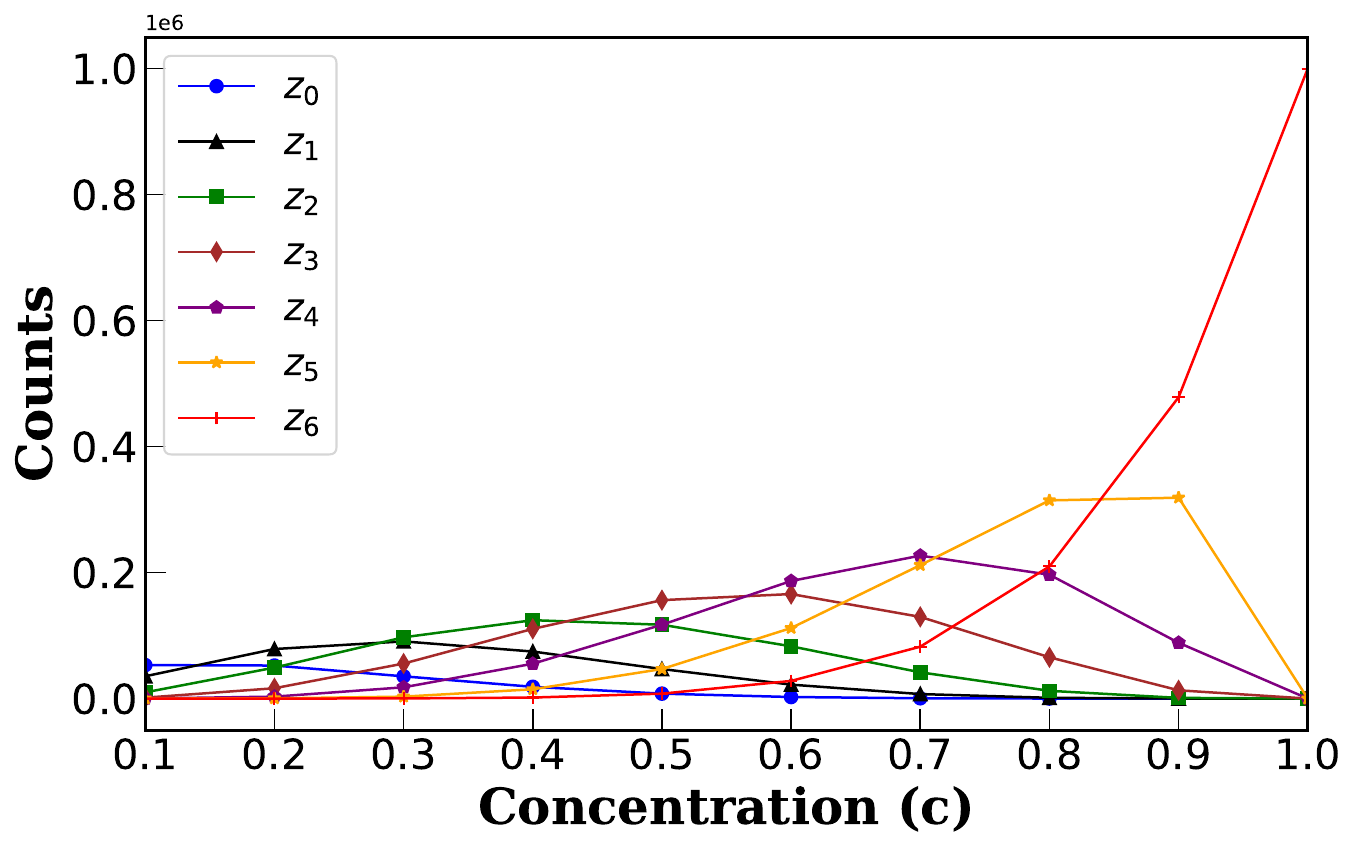}
    \caption{Distribution of different coordinated sites on varying the concentration $c$ on cubic lattice with linear dimension $N=200$, averaged over 10 000 configurations.}
    \label{fig:i8}
\end{figure}
We now study a dilute cubic system where $z=6$. There are no exact results available for the cubic system, so one has to rely on numerics alone. Fortunately, RFIM in the cubic system has been studied with great numerical accuracy \cite{perkovic1999disorder,d19973d,vives2004avalanche,wu2006numerical,picco2015diluted,fytas2016efficient}. Proceeding in the same manner as above, we start by diluting the system with probability $(1-c)$.
As we dilute the system, it becomes a mixture of different coordination numbers \( z=6, 5, 4, 3, 2, 1 \) and \( z_0 \) (isolated sites). We examine how the distribution of the different coordinated sites changes as the dilution is varied. This is shown in Fig.~\ref{fig:i8}. 
We find similar behavior in the distribution of the different $z$ coordinated sites and the usual shift of their peak towards the left with decreasing $z$ on lowering the concentration $c$. 

\begin{figure}[htb]
    \centering
   \includegraphics[width=8cm,height=4.5cm]{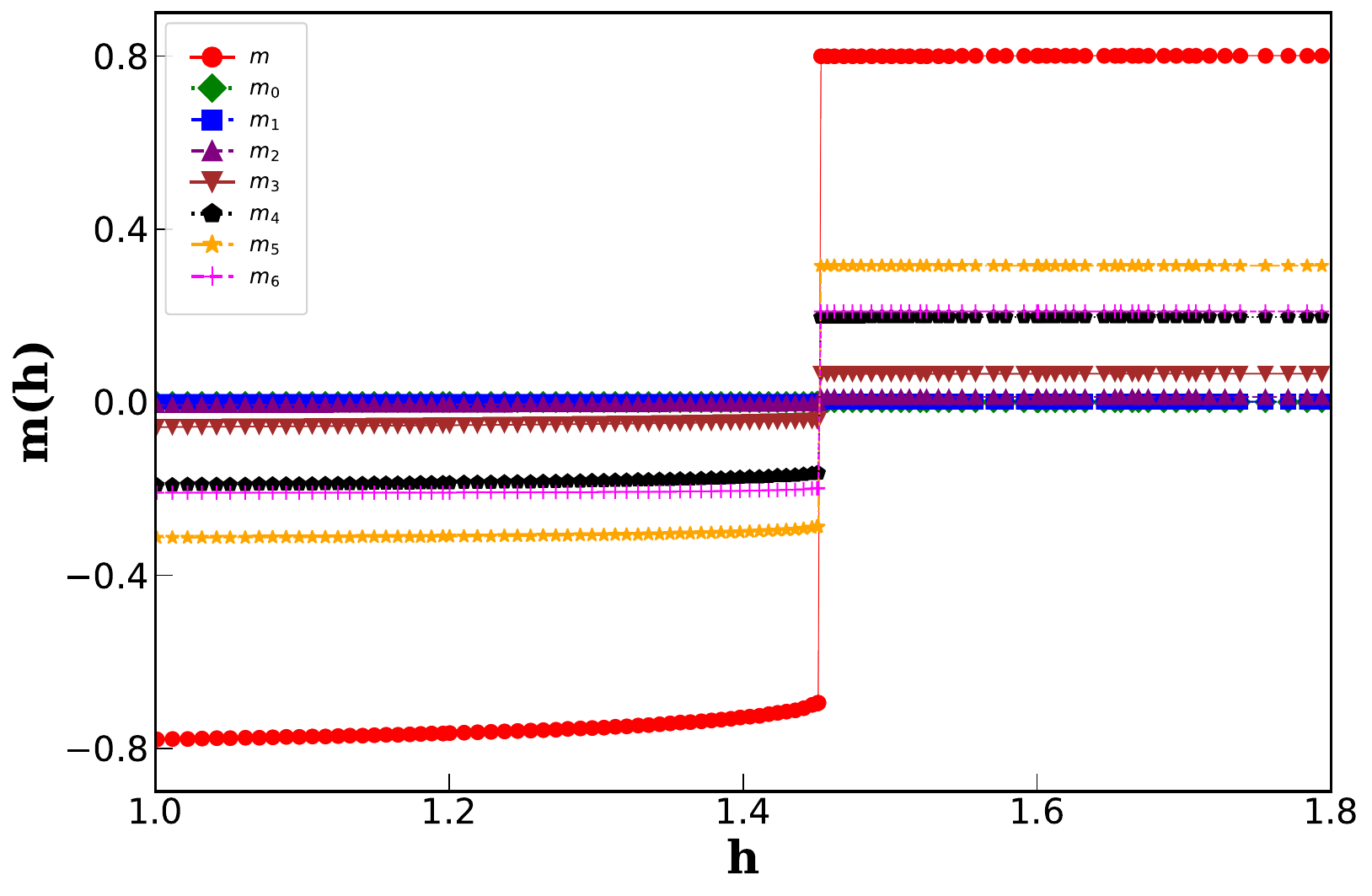}
    \caption{$m(h)$ vs $h$ on diluted cubic lattice showing jump discontinuities for $ z = 6, 5, 4, 3, 2,$ and $1$ sites. At $c=0.8$ (from  Fig.~\ref{fig:i8}), $z=5$ sublattice  has the highest number of sites  and hence contributes the maximum to the magnetization. The result is obtained for system size $N= 200^3$ and $\sigma = 1.2$.
}
    \label{fig:i9}
\end{figure}

We now consider $c=0.8$ for a cubic lattice of size $N=200^3$ and $\sigma=1.2$, where the jump discontinuity occurs. 
We analyze the magnetization curve for the entire cubic lattice ($m(h)$ vs $h$) and also separate contributions from each $z=6,5,4,3,2,1,$ and $0$ coordinated sites denoted as $m_6, m_5, m_4, m_3, m_2, m_1$, and $m_0$, respectively, denoted in Fig.~\ref{fig:i9}. 
We observe that the contribution of $m_5$ is more than $m_6$ for the chosen parameters. This is expected as, from Fig.~\ref{fig:i8} when $c=0.8$, the concentration of $z_5$ sites is more than that of $z_6$. 
\begin{figure}[h]
    \centering
    \includegraphics[width=8.8cm, height=9.1cm]{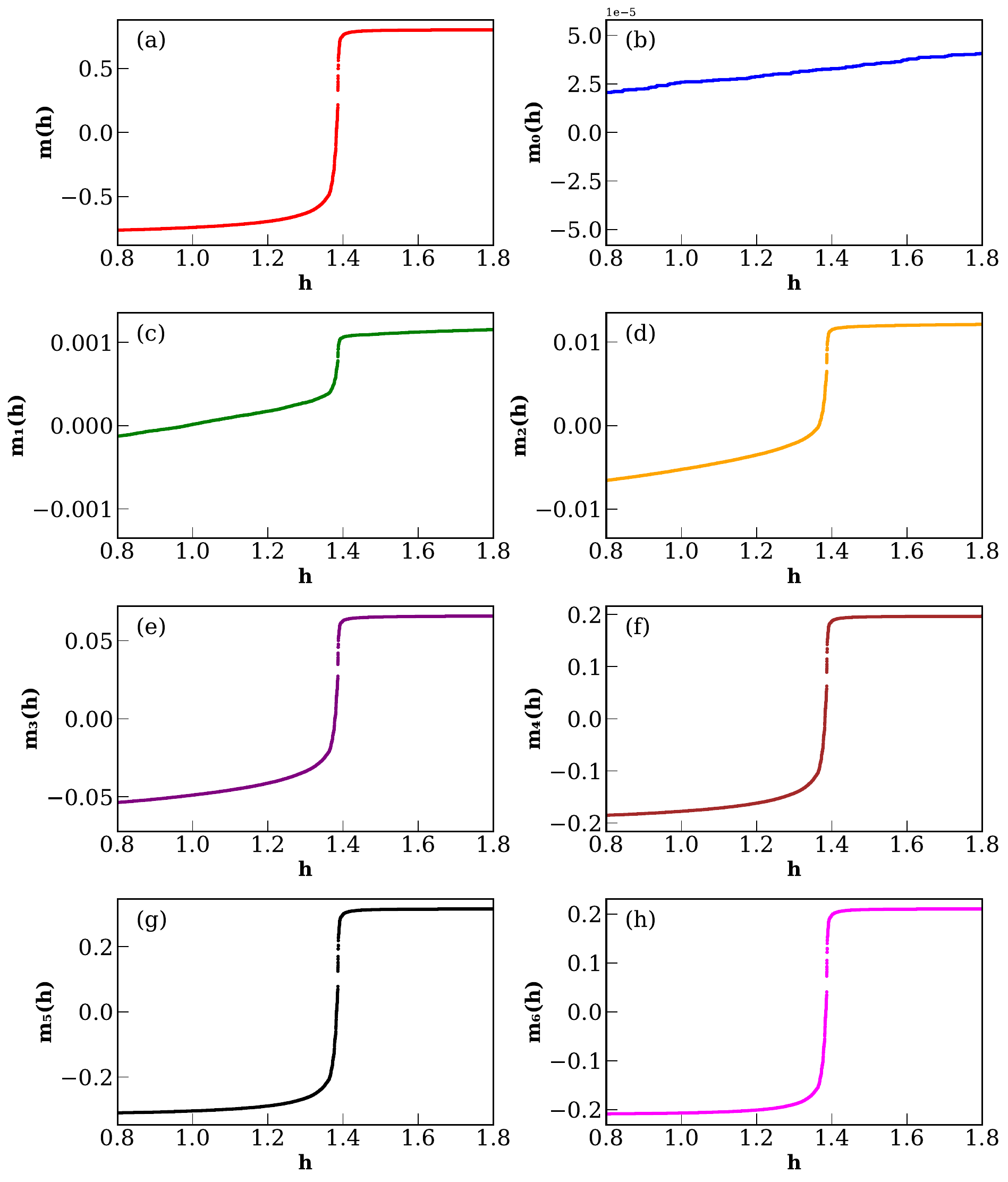}
    \caption{Plots of $m(h)$ vs $h$ at $\sigma \approx \sigma_c$ for system size $N=200^3$ and $c=0.8$ on dilute cubic lattice showing several breaks in magnetization consistent to the random graph simulation results, with the exception of $z=0$ sites. Results are shown for the whole sites in (a), and for $z=0,1,2,3,4,5,$ and $6$ sublattices in (b)-(h), respectively.
}
    \label{fig:i10}
\end{figure}

\begin{figure}[htb]
    \centering
   \includegraphics[width=8cm, height=5cm]{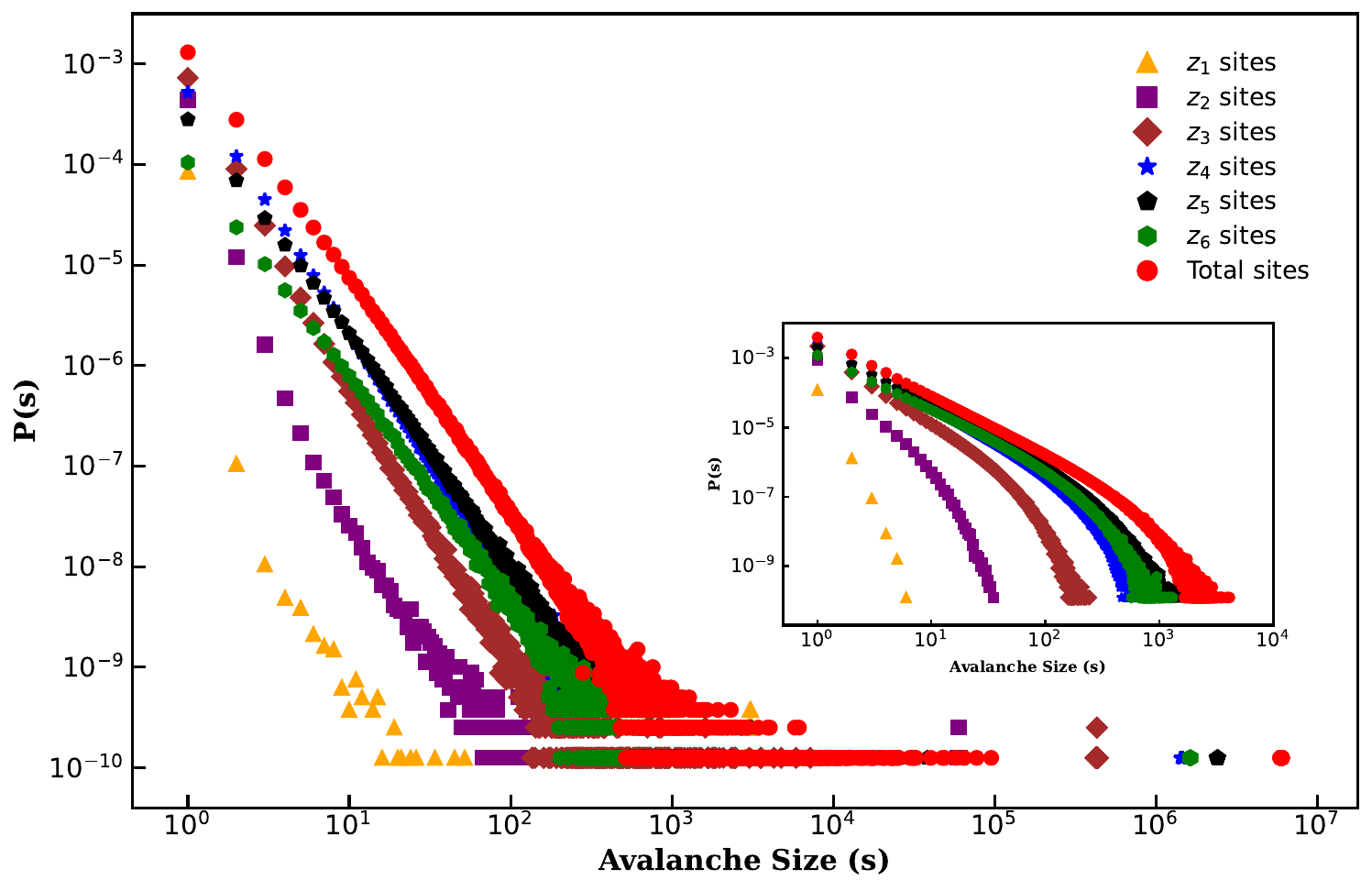}
    \caption{Integrated avalanche size distribution for the whole lattice and $z=1,2,3,4,5$ and 6 sublattices. The simulation data is obtained on cubic lattice with total $N= 200^3$ sites, $c = 0.8$, and $\sigma=1.2$ for 100 independent configurations. The distribution is shown on a log-log plot.
}
    \label{fig:i11a}
\end{figure}

 The data confirm that the jump discontinuity is not only restricted to \( z_6, z_5, z_4 \) sites rather other $z$ sites also contribute to the jump discontinuity. In fact, the different $z$ coordinated sublattices superimposed to give the total jump discontinuity in the lattice at the same value of external field $h_{crit}$. This is asserted on choosing different $\sigma=1.5$ close to $\sigma_c$ as shown in Figure~\ref{fig:i10}.
 
  The avalanche size distribution for the entire cubic lattice and $z=1,2,3,4,5,$ and 6 sublattices are shown in Fig.~\ref{fig:i11a}. The data is obtained by averaging over 100 different configurations on a cubic lattice with $N=200^3$ sites and $c=0.8$. Distribution of avalanche size includes all the avalanches when the $h$ is varied from $-\infty$ to $+\infty$. The main plot data for $\sigma=1.2$ exhibits power law as predicted. The results shown in the inset for $\sigma=2>\sigma_c(N)$ follow a linear trend over few decades before bending down in the log-log plot. The calculated value of power law exponent $\tau$ [$P(s)\approx s^{-\tau}$] in the linear regime for the whole cubic lattice and, $z=1,2,3,4,5,$ and 6 sublattices are $2.37\pm0.02$, $1.03\pm0.25$, $1.24\pm0.06$, $1.68\pm0.04$, $2.07\pm0.02$, $2.12\pm0.02$, and $1.89\pm0.03$, respectively. 
 
\section{Conclusion}
 To summarize, our study finds that the jump discontinuity is not limited to $z\geq4$ coordinated sites\cite{dhar1997zero}-it also arises for $z\leq3$ coordinated sites with the exception of $z=0$ sites for both the Bethe lattice and cubic lattice.  Our simulation results on random graph match exactly with the analytical results obtained on the Bethe lattice. The observed jump discontinuity is the superposition of jump discontinuities from $1\le z\leq4$ coordinated sites and occurs at the same value of $h=h_{crit}$. 
We observe that sites with higher concentration for a given value of $c$ have the largest contribution to the jump discontinuity, although sites with $z\geq3$ trigger the largest jump. We establish similar results on the cubic lattice, where analytical solution is elusive.

The task of finding the critical point is tedious and difficult
because the jump discontinuity is often accompanied by several breaks in the magnetization curve on approaching the critical point. The understanding of these fluctuations has been made possible to some extent by the presence of analytical results on the Bethe lattice. However, in the present study, we have only considered disordered systems with dilution.

\par
Interestingly, in dilute RFIM, there is an intricate interplay of vacant sites and onsite random field which competes
with the external field to restrict spin flips, thus fracturing avalanches, whereas the dilution breaks the connectivity between the sites. 
Similar jumps in magnetization curve are also observed in undiluted disordered systems for disorder strength $\sigma<\sigma_c$ in the context of RFIM whenever few $z\ge4$ sites are present. As shown in Fig.~\ref{fig:i4} and Fig.~\ref{fig:i9}, we also find jump discontinuity for $z\leq3$ sites (except for $z=0$ sites). This is because, after dilution, the sites with $z\leq3$ may still have connection with the sites having $z\geq4$ for both Bethe lattice and cubic lattice. As in the case of dilute RFIM with $z=4$, the presence of even a small fraction of $z=4$ sites with $c>0.56$ causes an avalanche \cite{diana2017criteria}. Also in Fig.~\ref{fig:i7} and Fig.~\ref{fig:i10}, at $\sigma \approx\sigma_c$, we find several breaks in magnetization. This is due to divergence in correlation length and the occurrence of avalanches in all scales on nearing the critical point, exhibited by the signature power-law distribution of the avalanche. Further, in the diluted case, there is reduced connectivity between sites which restricts the propagation of avalanches, which in turn inhibits jumps, resulting in several breaks in magnetization.

The jump discontinuity observed in the slowly driven non-equilibrium zero temperature RFIM systems could be related to the experimentally manifested abrupt phenomena in disordered systems. Recent experimental realization of 1D RFIM on an artificial spin ice system confirms avalanche dynamics \cite{bingham2021experimental}. In addition, previous studies have also shown realization of RFIM in physical systems viz. diluted antiferromagnets, diluted dipolar magnets, ferroelectric single crystals, and molecular magnets \cite{fishman1979random,hill1993magnetic,schechter2008liho,miga2009three,westphal1992diffuse,millis2010pure,xu2015barkhausen}.  In athermal martensitic transformation of binary alloys,  jump discontinuities are observed as strain and acoustic emission bursts\cite{sreekala2003acoustic,sreekala2004precursors,vives1994distributions,vives1995statistics}. Similarly, in ferromagnetic materials they are observed as discrete breaks in the magnetization called Barkhausen noise and as a discontinuous hysteresis loop in magnetic rocks \cite{kharwanglang,sethna1993hysteresis}. Real magnetic rocks or binary alloys are not inherently perfect and are always accompanied by some imperfections in the form of vacant sites or structural defects (compositional defects), which fractures the connectivity of the atoms resulting in fragmented atomic coordination. These types of materials perfectly align with the dilute RFIM studied here. Triggering sites initiates the nucleation event (or local instabilities) and, depending on the local environment, this nucleation event may turn into a localized event or jump discontinuities \cite{dahmen1994disorder,vives1995statistics}.

Understanding jump discontinuity is important in view of the fact that the observed avalanche as a response to a driving field could serve as a desirable or undesirable trait depending on the objective of the theoretical study or experiments on certain materials.
We believe our work although confined to studies of disordered systems in dilute form has relevance in the larger understanding of propagation of avalanches as we find the jumps are overlapping of the jumps from different  $z$ sublattices and the maximum contribution comes from the sublattice with the largest concentration. However, when concentration of $z_2$ is the largest in the window $c\to[0.556,0.6]$, most contributions to the $m(h)$ curve come from the $z_2$ sublattice, although the triggering sites for the jump are $z\ge3$ on the Bethe lattice.

\section*{Acknowledgements}
A.D. and D.T. acknowledge the Department of Science
and Technology, Anusandhan National Research Foundation,ANRF (then DST-SERB), Government of India, for financial support under POWER Grant No. SPG/2022/000678.
D.T. also acknowledges partial support from ANRF under the
Partnerships for Accelerated Innovation and Research (PAIR)
for the project JNU-PAIR network on Science for Sustainable
Future, File No. ANRF/PAIR/2025/000029/PAIR.

\section*{Data Availability}
The data that support the findings of this article are not
publicly available. The data are available from the authors
upon reasonable request.

\bibliography{drfim}

\begin{thebibliography}{49}%
\makeatletter
\providecommand \@ifxundefined [1]{%
 \@ifx{#1\undefined}
}%
\providecommand \@ifnum [1]{%
 \ifnum #1\expandafter \@firstoftwo
 \else \expandafter \@secondoftwo
 \fi
}%
\providecommand \@ifx [1]{%
 \ifx #1\expandafter \@firstoftwo
 \else \expandafter \@secondoftwo
 \fi
}%
\providecommand \natexlab [1]{#1}%
\providecommand \enquote  [1]{``#1''}%
\providecommand \bibnamefont  [1]{#1}%
\providecommand \bibfnamefont [1]{#1}%
\providecommand \citenamefont [1]{#1}%
\providecommand \href@noop [0]{\@secondoftwo}%
\providecommand \href [0]{\begingroup \@sanitize@url \@href}%
\providecommand \@href[1]{\@@startlink{#1}\@@href}%
\providecommand \@@href[1]{\endgroup#1\@@endlink}%
\providecommand \@sanitize@url [0]{\catcode `\\12\catcode `\$12\catcode
  `\&12\catcode `\#12\catcode `\^12\catcode `\_12\catcode `\%12\relax}%
\providecommand \@@startlink[1]{}%
\providecommand \@@endlink[0]{}%
\providecommand \url  [0]{\begingroup\@sanitize@url \@url }%
\providecommand \@url [1]{\endgroup\@href {#1}{\urlprefix }}%
\providecommand \urlprefix  [0]{URL }%
\providecommand \Eprint [0]{\href }%
\providecommand \doibase [0]{https://doi.org/}%
\providecommand \selectlanguage [0]{\@gobble}%
\providecommand \bibinfo  [0]{\@secondoftwo}%
\providecommand \bibfield  [0]{\@secondoftwo}%
\providecommand \translation [1]{[#1]}%
\providecommand \BibitemOpen [0]{}%
\providecommand \bibitemStop [0]{}%
\providecommand \bibitemNoStop [0]{.\EOS\space}%
\providecommand \EOS [0]{\spacefactor3000\relax}%
\providecommand \BibitemShut  [1]{\csname bibitem#1\endcsname}%
\let\auto@bib@innerbib\@empty
\bibitem [{\citenamefont {Sethna}\ \emph {et~al.}(1993)\citenamefont {Sethna},
  \citenamefont {Dahmen}, \citenamefont {Kartha}, \citenamefont {Krumhansl},
  \citenamefont {Roberts},\ and\ \citenamefont {Shore}}]{sethna1993hysteresis}%
  \BibitemOpen
  \bibfield  {author} {\bibinfo {author} {\bibfnamefont {J.~P.}\ \bibnamefont
  {Sethna}}, \bibinfo {author} {\bibfnamefont {K.}~\bibnamefont {Dahmen}},
  \bibinfo {author} {\bibfnamefont {S.}~\bibnamefont {Kartha}}, \bibinfo
  {author} {\bibfnamefont {J.~A.}\ \bibnamefont {Krumhansl}}, \bibinfo {author}
  {\bibfnamefont {B.~W.}\ \bibnamefont {Roberts}},\ and\ \bibinfo {author}
  {\bibfnamefont {J.~D.}\ \bibnamefont {Shore}},\ }\href@noop {} {\bibfield
  {journal} {\bibinfo  {journal} {Physical Review Letters}\ }\textbf {\bibinfo
  {volume} {70}},\ \bibinfo {pages} {3347} (\bibinfo {year}
  {1993})}\BibitemShut {NoStop}%
\bibitem [{\citenamefont {Wohlman}\ and\ \citenamefont
  {Domb}(1984)}]{wohlman1984random}%
  \BibitemOpen
  \bibfield  {author} {\bibinfo {author} {\bibfnamefont {O.~E.}\ \bibnamefont
  {Wohlman}}\ and\ \bibinfo {author} {\bibfnamefont {C.}~\bibnamefont {Domb}},\
  }\href@noop {} {\bibfield  {journal} {\bibinfo  {journal} {Journal of Physics
  A: Mathematical and General}\ }\textbf {\bibinfo {volume} {17}},\ \bibinfo
  {pages} {2247} (\bibinfo {year} {1984})}\BibitemShut {NoStop}%
\bibitem [{\citenamefont {Nattermann}(1997)}]{nattermann1997}%
  \BibitemOpen
  \bibfield  {author} {\bibinfo {author} {\bibfnamefont {T.}~\bibnamefont
  {Nattermann}},\ }in\ \href@noop {} {\emph {\bibinfo {booktitle} {Spin Glasses
  and Random Fields}}},\ \bibinfo {editor} {edited by\ \bibinfo {editor}
  {\bibfnamefont {A.~P.}\ \bibnamefont {Young}}}\ (\bibinfo  {publisher} {World
  Scientific},\ \bibinfo {address} {Singapore},\ \bibinfo {year} {1997})\ pp.\
  \bibinfo {pages} {277--298}\BibitemShut {NoStop}%
\bibitem [{\citenamefont {Sabhapandit}\ \emph {et~al.}(2000)\citenamefont
  {Sabhapandit}, \citenamefont {Shukla},\ and\ \citenamefont
  {Dhar}}]{sabhapandit2000distribution}%
  \BibitemOpen
  \bibfield  {author} {\bibinfo {author} {\bibfnamefont {S.}~\bibnamefont
  {Sabhapandit}}, \bibinfo {author} {\bibfnamefont {P.}~\bibnamefont
  {Shukla}},\ and\ \bibinfo {author} {\bibfnamefont {D.}~\bibnamefont {Dhar}},\
  }\href@noop {} {\bibfield  {journal} {\bibinfo  {journal} {Journal of
  Statistical Physics}\ }\textbf {\bibinfo {volume} {98}},\ \bibinfo {pages}
  {103} (\bibinfo {year} {2000})}\BibitemShut {NoStop}%
\bibitem [{\citenamefont {Sethna}\ \emph {et~al.}(2001)\citenamefont {Sethna},
  \citenamefont {Dahmen},\ and\ \citenamefont {Myers}}]{sethna2001crackling}%
  \BibitemOpen
  \bibfield  {author} {\bibinfo {author} {\bibfnamefont {J.~P.}\ \bibnamefont
  {Sethna}}, \bibinfo {author} {\bibfnamefont {K.~A.}\ \bibnamefont {Dahmen}},\
  and\ \bibinfo {author} {\bibfnamefont {C.~R.}\ \bibnamefont {Myers}},\
  }\href@noop {} {\bibfield  {journal} {\bibinfo  {journal} {Nature}\ }\textbf
  {\bibinfo {volume} {410}},\ \bibinfo {pages} {242} (\bibinfo {year}
  {2001})}\BibitemShut {NoStop}%
\bibitem [{\citenamefont {Vives}\ and\ \citenamefont
  {Planes}(2000)}]{vives2000hysteresis}%
  \BibitemOpen
  \bibfield  {author} {\bibinfo {author} {\bibfnamefont {E.}~\bibnamefont
  {Vives}}\ and\ \bibinfo {author} {\bibfnamefont {A.}~\bibnamefont {Planes}},\
  }\href@noop {} {\bibfield  {journal} {\bibinfo  {journal} {Journal of
  magnetism and magnetic materials}\ }\textbf {\bibinfo {volume} {221}},\
  \bibinfo {pages} {164} (\bibinfo {year} {2000})}\BibitemShut {NoStop}%
\bibitem [{\citenamefont {Jani{\'c}evi{\'c}}\ \emph {et~al.}(2017)\citenamefont
  {Jani{\'c}evi{\'c}}, \citenamefont {Mijatovi{\'c}},\ and\ \citenamefont
  {Spasojevi{\'c}}}]{janicevic2017critical}%
  \BibitemOpen
  \bibfield  {author} {\bibinfo {author} {\bibfnamefont {S.}~\bibnamefont
  {Jani{\'c}evi{\'c}}}, \bibinfo {author} {\bibfnamefont {S.}~\bibnamefont
  {Mijatovi{\'c}}},\ and\ \bibinfo {author} {\bibfnamefont {D.}~\bibnamefont
  {Spasojevi{\'c}}},\ }\href@noop {} {\bibfield  {journal} {\bibinfo  {journal}
  {Physical Review E}\ }\textbf {\bibinfo {volume} {95}},\ \bibinfo {pages}
  {042131} (\bibinfo {year} {2017})}\BibitemShut {NoStop}%
\bibitem [{\citenamefont {Spasojevi{\'c}}\ \emph {et~al.}(2018)\citenamefont
  {Spasojevi{\'c}}, \citenamefont {Mijatovi{\'c}}, \citenamefont
  {Navas-Portella},\ and\ \citenamefont {Vives}}]{spasojevic2018crossover}%
  \BibitemOpen
  \bibfield  {author} {\bibinfo {author} {\bibfnamefont {D.}~\bibnamefont
  {Spasojevi{\'c}}}, \bibinfo {author} {\bibfnamefont {S.}~\bibnamefont
  {Mijatovi{\'c}}}, \bibinfo {author} {\bibfnamefont {V.}~\bibnamefont
  {Navas-Portella}},\ and\ \bibinfo {author} {\bibfnamefont {E.}~\bibnamefont
  {Vives}},\ }\href@noop {} {\bibfield  {journal} {\bibinfo  {journal}
  {Physical Review E}\ }\textbf {\bibinfo {volume} {97}},\ \bibinfo {pages}
  {012109} (\bibinfo {year} {2018})}\BibitemShut {NoStop}%
\bibitem [{\citenamefont {Mijatovi{\'c}}\ \emph {et~al.}(2021)\citenamefont
  {Mijatovi{\'c}}, \citenamefont {Jovkovi{\'c}},\ and\ \citenamefont
  {Spasojevi{\'c}}}]{mijatovic2021nonequilibrium}%
  \BibitemOpen
  \bibfield  {author} {\bibinfo {author} {\bibfnamefont {S.}~\bibnamefont
  {Mijatovi{\'c}}}, \bibinfo {author} {\bibfnamefont {D.}~\bibnamefont
  {Jovkovi{\'c}}},\ and\ \bibinfo {author} {\bibfnamefont {D.}~\bibnamefont
  {Spasojevi{\'c}}},\ }\href@noop {} {\bibfield  {journal} {\bibinfo  {journal}
  {Physical Review E}\ }\textbf {\bibinfo {volume} {103}},\ \bibinfo {pages}
  {032147} (\bibinfo {year} {2021})}\BibitemShut {NoStop}%
\bibitem [{\citenamefont {Bingham}\ \emph {et~al.}(2021)\citenamefont
  {Bingham}, \citenamefont {Rooke}, \citenamefont {Park}, \citenamefont
  {Simon}, \citenamefont {Zhu}, \citenamefont {Zhang}, \citenamefont {Batley},
  \citenamefont {Watts}, \citenamefont {Leighton}, \citenamefont {Dahmen} \emph
  {et~al.}}]{bingham2021experimental}%
  \BibitemOpen
  \bibfield  {author} {\bibinfo {author} {\bibfnamefont {N.}~\bibnamefont
  {Bingham}}, \bibinfo {author} {\bibfnamefont {S.}~\bibnamefont {Rooke}},
  \bibinfo {author} {\bibfnamefont {J.}~\bibnamefont {Park}}, \bibinfo {author}
  {\bibfnamefont {A.}~\bibnamefont {Simon}}, \bibinfo {author} {\bibfnamefont
  {W.}~\bibnamefont {Zhu}}, \bibinfo {author} {\bibfnamefont {X.}~\bibnamefont
  {Zhang}}, \bibinfo {author} {\bibfnamefont {J.}~\bibnamefont {Batley}},
  \bibinfo {author} {\bibfnamefont {J.}~\bibnamefont {Watts}}, \bibinfo
  {author} {\bibfnamefont {C.}~\bibnamefont {Leighton}}, \bibinfo {author}
  {\bibfnamefont {K.}~\bibnamefont {Dahmen}}, \emph {et~al.},\ }\href@noop {}
  {\bibfield  {journal} {\bibinfo  {journal} {Physical review letters}\
  }\textbf {\bibinfo {volume} {127}},\ \bibinfo {pages} {207203} (\bibinfo
  {year} {2021})}\BibitemShut {NoStop}%
\bibitem [{\citenamefont {Rosinberg}\ \emph {et~al.}(2009)\citenamefont
  {Rosinberg}, \citenamefont {Tarjus},\ and\ \citenamefont
  {Pérez-Reche}}]{rosinberg2009t}%
  \BibitemOpen
  \bibfield  {author} {\bibinfo {author} {\bibfnamefont {M.~L.}\ \bibnamefont
  {Rosinberg}}, \bibinfo {author} {\bibfnamefont {G.}~\bibnamefont {Tarjus}},\
  and\ \bibinfo {author} {\bibfnamefont {F.~J.}\ \bibnamefont {Pérez-Reche}},\
  }\href@noop {} {\bibfield  {journal} {\bibinfo  {journal} {Journal of
  Statistical Mechanics: Theory and Experiment}\ }\textbf {\bibinfo {volume}
  {2009}},\ \bibinfo {pages} {P03003} (\bibinfo {year} {2009})}\BibitemShut
  {NoStop}%
\bibitem [{\citenamefont {Illa}\ and\ \citenamefont
  {Vives}(2006)}]{illa2006diluted}%
  \BibitemOpen
  \bibfield  {author} {\bibinfo {author} {\bibfnamefont {X.}~\bibnamefont
  {Illa}}\ and\ \bibinfo {author} {\bibfnamefont {E.}~\bibnamefont {Vives}},\
  }\href@noop {} {\bibfield  {journal} {\bibinfo  {journal} {Physical Review
  B—Condensed Matter and Materials Physics}\ }\textbf {\bibinfo {volume}
  {74}},\ \bibinfo {pages} {104409} (\bibinfo {year} {2006})}\BibitemShut
  {NoStop}%
\bibitem [{\citenamefont {Aharony}\ \emph {et~al.}(1985)\citenamefont
  {Aharony}, \citenamefont {Harris},\ and\ \citenamefont
  {Meir}}]{aharony1985dilute}%
  \BibitemOpen
  \bibfield  {author} {\bibinfo {author} {\bibfnamefont {A.}~\bibnamefont
  {Aharony}}, \bibinfo {author} {\bibfnamefont {A.~B.}\ \bibnamefont
  {Harris}},\ and\ \bibinfo {author} {\bibfnamefont {Y.}~\bibnamefont {Meir}},\
  }\href@noop {} {\bibfield  {journal} {\bibinfo  {journal} {Physical Review
  B}\ }\textbf {\bibinfo {volume} {32}},\ \bibinfo {pages} {3203} (\bibinfo
  {year} {1985})}\BibitemShut {NoStop}%
\bibitem [{\citenamefont {Dhar}\ \emph {et~al.}(1997)\citenamefont {Dhar},
  \citenamefont {Shukla},\ and\ \citenamefont {Sethna}}]{dhar1997zero}%
  \BibitemOpen
  \bibfield  {author} {\bibinfo {author} {\bibfnamefont {D.}~\bibnamefont
  {Dhar}}, \bibinfo {author} {\bibfnamefont {P.}~\bibnamefont {Shukla}},\ and\
  \bibinfo {author} {\bibfnamefont {J.~P.}\ \bibnamefont {Sethna}},\
  }\href@noop {} {\bibfield  {journal} {\bibinfo  {journal} {Journal of Physics
  A: Mathematical and General}\ }\textbf {\bibinfo {volume} {30}},\ \bibinfo
  {pages} {5259} (\bibinfo {year} {1997})}\BibitemShut {NoStop}%
\bibitem [{\citenamefont {Shukla}(1996)}]{shukla1996hysteresis}%
  \BibitemOpen
  \bibfield  {author} {\bibinfo {author} {\bibfnamefont {P.}~\bibnamefont
  {Shukla}},\ }\href@noop {} {\bibfield  {journal} {\bibinfo  {journal}
  {Progress of Theoretical Physics}\ }\textbf {\bibinfo {volume} {96}},\
  \bibinfo {pages} {69} (\bibinfo {year} {1996})}\BibitemShut {NoStop}%
\bibitem [{\citenamefont {Sabhapandit}(2004)}]{sabhapandit2004absence}%
  \BibitemOpen
  \bibfield  {author} {\bibinfo {author} {\bibfnamefont {S.}~\bibnamefont
  {Sabhapandit}},\ }\href@noop {} {\bibfield  {journal} {\bibinfo  {journal}
  {Physical Review B—Condensed Matter and Materials Physics}\ }\textbf
  {\bibinfo {volume} {70}},\ \bibinfo {pages} {224401} (\bibinfo {year}
  {2004})}\BibitemShut {NoStop}%
\bibitem [{\citenamefont {Bupathy}\ \emph {et~al.}(2017)\citenamefont
  {Bupathy}, \citenamefont {Kumar}, \citenamefont {Banerjee},\ and\
  \citenamefont {Puri}}]{bupathy2017random}%
  \BibitemOpen
  \bibfield  {author} {\bibinfo {author} {\bibfnamefont {A.}~\bibnamefont
  {Bupathy}}, \bibinfo {author} {\bibfnamefont {M.}~\bibnamefont {Kumar}},
  \bibinfo {author} {\bibfnamefont {V.}~\bibnamefont {Banerjee}},\ and\
  \bibinfo {author} {\bibfnamefont {S.}~\bibnamefont {Puri}},\ }\href@noop {}
  {\bibfield  {journal} {\bibinfo  {journal} {Journal of Physics: Conference
  Series}\ }\textbf {\bibinfo {volume} {905}},\ \bibinfo {pages} {012025}
  (\bibinfo {year} {2017})}\BibitemShut {NoStop}%
\bibitem [{\citenamefont {Kharwanlang}\ and\ \citenamefont
  {Shukla}(2012)}]{kharwanglang}%
  \BibitemOpen
  \bibfield  {author} {\bibinfo {author} {\bibfnamefont {R.~S.}\ \bibnamefont
  {Kharwanlang}}\ and\ \bibinfo {author} {\bibfnamefont {P.}~\bibnamefont
  {Shukla}},\ }\href@noop {} {\bibfield  {journal} {\bibinfo  {journal}
  {Physical Review E}\ }\textbf {\bibinfo {volume} {85}} (\bibinfo {year}
  {2012})}\BibitemShut {NoStop}%
\bibitem [{\citenamefont {Liu}\ and\ \citenamefont
  {Dahmen}(2009)}]{liu2009unexpected}%
  \BibitemOpen
  \bibfield  {author} {\bibinfo {author} {\bibfnamefont {Y.}~\bibnamefont
  {Liu}}\ and\ \bibinfo {author} {\bibfnamefont {K.~A.}\ \bibnamefont
  {Dahmen}},\ }\href@noop {} {\bibfield  {journal} {\bibinfo  {journal} {Phys.
  Rev. E}\ }\textbf {\bibinfo {volume} {79}},\ \bibinfo {pages} {061124}
  (\bibinfo {year} {2009})}\BibitemShut {NoStop}%
\bibitem [{\citenamefont {Imry}\ and\ \citenamefont
  {Ma}(1975)}]{imry1975random}%
  \BibitemOpen
  \bibfield  {author} {\bibinfo {author} {\bibfnamefont {Y.}~\bibnamefont
  {Imry}}\ and\ \bibinfo {author} {\bibfnamefont {S.-k.}\ \bibnamefont {Ma}},\
  }\href@noop {} {\bibfield  {journal} {\bibinfo  {journal} {Physical Review
  Letters}\ }\textbf {\bibinfo {volume} {35}},\ \bibinfo {pages} {1399}
  (\bibinfo {year} {1975})}\BibitemShut {NoStop}%
\bibitem [{\citenamefont {Shukla}(2000)}]{shukla2000exact}%
  \BibitemOpen
  \bibfield  {author} {\bibinfo {author} {\bibfnamefont {P.}~\bibnamefont
  {Shukla}},\ }\href@noop {} {\bibfield  {journal} {\bibinfo  {journal} {Phys.
  Rev. E}\ }\textbf {\bibinfo {volume} {62}},\ \bibinfo {pages} {4725}
  (\bibinfo {year} {2000})}\BibitemShut {NoStop}%
\bibitem [{\citenamefont {Imbrie}(1984)}]{imbrie1984lower}%
  \BibitemOpen
  \bibfield  {author} {\bibinfo {author} {\bibfnamefont {J.~Z.}\ \bibnamefont
  {Imbrie}},\ }\href@noop {} {\bibfield  {journal} {\bibinfo  {journal}
  {Physical Review Letters}\ }\textbf {\bibinfo {volume} {53}},\ \bibinfo
  {pages} {1747} (\bibinfo {year} {1984})}\BibitemShut {NoStop}%
\bibitem [{\citenamefont {Shukla}\ and\ \citenamefont
  {Thongjaomayum}(2016)}]{diana2016hysteresis}%
  \BibitemOpen
  \bibfield  {author} {\bibinfo {author} {\bibfnamefont {P.}~\bibnamefont
  {Shukla}}\ and\ \bibinfo {author} {\bibfnamefont {D.}~\bibnamefont
  {Thongjaomayum}},\ }\href@noop {} {\bibfield  {journal} {\bibinfo  {journal}
  {Journal of Physics A: Mathematical and Theoretical}\ }\textbf {\bibinfo
  {volume} {49}},\ \bibinfo {pages} {235001} (\bibinfo {year}
  {2016})}\BibitemShut {NoStop}%
\bibitem [{\citenamefont {Shukla}\ and\ \citenamefont
  {Thongjaomayum}(2017)}]{diana2017criteria}%
  \BibitemOpen
  \bibfield  {author} {\bibinfo {author} {\bibfnamefont {P.}~\bibnamefont
  {Shukla}}\ and\ \bibinfo {author} {\bibfnamefont {D.}~\bibnamefont
  {Thongjaomayum}},\ }\href@noop {} {\bibfield  {journal} {\bibinfo  {journal}
  {Physical Review E}\ }\textbf {\bibinfo {volume} {95}},\ \bibinfo {pages}
  {042109} (\bibinfo {year} {2017})}\BibitemShut {NoStop}%
\bibitem [{\citenamefont {Sabhapandit}\ \emph {et~al.}(2002)\citenamefont
  {Sabhapandit}, \citenamefont {Dhar},\ and\ \citenamefont
  {Shukla}}]{sabhapandit2002hysteresis}%
  \BibitemOpen
  \bibfield  {author} {\bibinfo {author} {\bibfnamefont {S.}~\bibnamefont
  {Sabhapandit}}, \bibinfo {author} {\bibfnamefont {D.}~\bibnamefont {Dhar}},\
  and\ \bibinfo {author} {\bibfnamefont {P.}~\bibnamefont {Shukla}},\
  }\href@noop {} {\bibfield  {journal} {\bibinfo  {journal} {Physical Review
  Letters}\ }\textbf {\bibinfo {volume} {88}},\ \bibinfo {pages} {197202}
  (\bibinfo {year} {2002})}\BibitemShut {NoStop}%
\bibitem [{\citenamefont {Spasojevi\ifmmode~\acute{c}\else \'{c}\fi{}}\ \emph
  {et~al.}(2011)\citenamefont {Spasojevi\ifmmode~\acute{c}\else \'{c}\fi{}},
  \citenamefont {Jani\ifmmode \acute{c}\else
  \'{c}\fi{}evi\ifmmode~\acute{c}\else \'{c}\fi{}},\ and\ \citenamefont
  {Kne\ifmmode \check{z}\else \v{z}\fi{}evi\ifmmode~\acute{c}\else
  \'{c}\fi{}}}]{spasojevic2011numerical}%
  \BibitemOpen
  \bibfield  {author} {\bibinfo {author} {\bibfnamefont {D.}~\bibnamefont
  {Spasojevi\ifmmode~\acute{c}\else \'{c}\fi{}}}, \bibinfo {author}
  {\bibfnamefont {S.}~\bibnamefont {Jani\ifmmode \acute{c}\else
  \'{c}\fi{}evi\ifmmode~\acute{c}\else \'{c}\fi{}}},\ and\ \bibinfo {author}
  {\bibfnamefont {M.}~\bibnamefont {Kne\ifmmode \check{z}\else
  \v{z}\fi{}evi\ifmmode~\acute{c}\else \'{c}\fi{}}},\ }\href@noop {} {\bibfield
   {journal} {\bibinfo  {journal} {Phys. Rev. Lett.}\ }\textbf {\bibinfo
  {volume} {106}},\ \bibinfo {pages} {175701} (\bibinfo {year}
  {2011})}\BibitemShut {NoStop}%
\bibitem [{\citenamefont {Thongjaomayum}\ and\ \citenamefont
  {Shukla}(2013)}]{diana2013effect}%
  \BibitemOpen
  \bibfield  {author} {\bibinfo {author} {\bibfnamefont {D.}~\bibnamefont
  {Thongjaomayum}}\ and\ \bibinfo {author} {\bibfnamefont {P.}~\bibnamefont
  {Shukla}},\ }\href@noop {} {\bibfield  {journal} {\bibinfo  {journal}
  {Physical Review E}\ }\textbf {\bibinfo {volume} {88}},\ \bibinfo {pages}
  {042138} (\bibinfo {year} {2013})}\BibitemShut {NoStop}%
\bibitem [{\citenamefont {Thongjaomayum}\ and\ \citenamefont
  {Shukla}(2019)}]{diana2019critical}%
  \BibitemOpen
  \bibfield  {author} {\bibinfo {author} {\bibfnamefont {D.}~\bibnamefont
  {Thongjaomayum}}\ and\ \bibinfo {author} {\bibfnamefont {P.}~\bibnamefont
  {Shukla}},\ }\href@noop {} {\bibfield  {journal} {\bibinfo  {journal}
  {Physical Review E}\ }\textbf {\bibinfo {volume} {99}},\ \bibinfo {pages}
  {062136} (\bibinfo {year} {2019})}\BibitemShut {NoStop}%
\bibitem [{\citenamefont {Kurbah}\ \emph {et~al.}(2015)\citenamefont {Kurbah},
  \citenamefont {Thongjaomayum},\ and\ \citenamefont
  {Shukla}}]{kurbah2015nonequilibrium}%
  \BibitemOpen
  \bibfield  {author} {\bibinfo {author} {\bibfnamefont {L.}~\bibnamefont
  {Kurbah}}, \bibinfo {author} {\bibfnamefont {D.}~\bibnamefont
  {Thongjaomayum}},\ and\ \bibinfo {author} {\bibfnamefont {P.}~\bibnamefont
  {Shukla}},\ }\href@noop {} {\bibfield  {journal} {\bibinfo  {journal}
  {Physical Review E}\ }\textbf {\bibinfo {volume} {91}},\ \bibinfo {pages}
  {012131} (\bibinfo {year} {2015})}\BibitemShut {NoStop}%
\bibitem [{\citenamefont {Fytas}\ and\ \citenamefont
  {Mart{\'\i}n-Mayor}(2013)}]{fytas2013universality}%
  \BibitemOpen
  \bibfield  {author} {\bibinfo {author} {\bibfnamefont {N.~G.}\ \bibnamefont
  {Fytas}}\ and\ \bibinfo {author} {\bibfnamefont {V.}~\bibnamefont
  {Mart{\'\i}n-Mayor}},\ }\href@noop {} {\bibfield  {journal} {\bibinfo
  {journal} {Physical Review Letters}\ }\textbf {\bibinfo {volume} {110}},\
  \bibinfo {pages} {227201} (\bibinfo {year} {2013})}\BibitemShut {NoStop}%
\bibitem [{\citenamefont {Glauber}(1963)}]{glauber1963time}%
  \BibitemOpen
  \bibfield  {author} {\bibinfo {author} {\bibfnamefont {R.~J.}\ \bibnamefont
  {Glauber}},\ }\href@noop {} {\bibfield  {journal} {\bibinfo  {journal}
  {Journal of Mathematical Physics}\ }\textbf {\bibinfo {volume} {4}},\
  \bibinfo {pages} {294} (\bibinfo {year} {1963})}\BibitemShut {NoStop}%
\bibitem [{\citenamefont {Perkovi{\'c}}\ \emph {et~al.}(1999)\citenamefont
  {Perkovi{\'c}}, \citenamefont {Dahmen},\ and\ \citenamefont
  {Sethna}}]{perkovic1999disorder}%
  \BibitemOpen
  \bibfield  {author} {\bibinfo {author} {\bibfnamefont {O.}~\bibnamefont
  {Perkovi{\'c}}}, \bibinfo {author} {\bibfnamefont {K.~A.}\ \bibnamefont
  {Dahmen}},\ and\ \bibinfo {author} {\bibfnamefont {J.~P.}\ \bibnamefont
  {Sethna}},\ }\href@noop {} {\bibfield  {journal} {\bibinfo  {journal}
  {Physical Review B}\ }\textbf {\bibinfo {volume} {59}},\ \bibinfo {pages}
  {6106} (\bibinfo {year} {1999})}\BibitemShut {NoStop}%
\bibitem [{\citenamefont {d'Auriac}\ and\ \citenamefont
  {Sourlas}(1997)}]{d19973d}%
  \BibitemOpen
  \bibfield  {author} {\bibinfo {author} {\bibfnamefont {J.-C.~A.}\
  \bibnamefont {d'Auriac}}\ and\ \bibinfo {author} {\bibfnamefont
  {N.}~\bibnamefont {Sourlas}},\ }\href@noop {} {\bibfield  {journal} {\bibinfo
   {journal} {Europhysics Letters}\ }\textbf {\bibinfo {volume} {39}},\
  \bibinfo {pages} {473} (\bibinfo {year} {1997})}\BibitemShut {NoStop}%
\bibitem [{\citenamefont {Vives}\ and\ \citenamefont
  {P{\'e}rez-Reche}(2004)}]{vives2004avalanche}%
  \BibitemOpen
  \bibfield  {author} {\bibinfo {author} {\bibfnamefont {E.}~\bibnamefont
  {Vives}}\ and\ \bibinfo {author} {\bibfnamefont {F.-J.}\ \bibnamefont
  {P{\'e}rez-Reche}},\ }\href@noop {} {\bibfield  {journal} {\bibinfo
  {journal} {Physica B: Condensed Matter}\ }\textbf {\bibinfo {volume} {343}},\
  \bibinfo {pages} {281} (\bibinfo {year} {2004})}\BibitemShut {NoStop}%
\bibitem [{\citenamefont {Wu}\ and\ \citenamefont
  {Machta}(2006)}]{wu2006numerical}%
  \BibitemOpen
  \bibfield  {author} {\bibinfo {author} {\bibfnamefont {Y.}~\bibnamefont
  {Wu}}\ and\ \bibinfo {author} {\bibfnamefont {J.}~\bibnamefont {Machta}},\
  }\href@noop {} {\bibfield  {journal} {\bibinfo  {journal} {Physical Review
  B—Condensed Matter and Materials Physics}\ }\textbf {\bibinfo {volume}
  {74}},\ \bibinfo {pages} {064418} (\bibinfo {year} {2006})}\BibitemShut
  {NoStop}%
\bibitem [{\citenamefont {Picco}\ and\ \citenamefont
  {Sourlas}(2015)}]{picco2015diluted}%
  \BibitemOpen
  \bibfield  {author} {\bibinfo {author} {\bibfnamefont {M.}~\bibnamefont
  {Picco}}\ and\ \bibinfo {author} {\bibfnamefont {N.}~\bibnamefont
  {Sourlas}},\ }\href@noop {} {\bibfield  {journal} {\bibinfo  {journal}
  {Europhysics Letters}\ }\textbf {\bibinfo {volume} {109}},\ \bibinfo {pages}
  {37001} (\bibinfo {year} {2015})}\BibitemShut {NoStop}%
\bibitem [{\citenamefont {Fytas}\ and\ \citenamefont
  {Mart{\'\i}n-Mayor}(2016)}]{fytas2016efficient}%
  \BibitemOpen
  \bibfield  {author} {\bibinfo {author} {\bibfnamefont {N.~G.}\ \bibnamefont
  {Fytas}}\ and\ \bibinfo {author} {\bibfnamefont {V.}~\bibnamefont
  {Mart{\'\i}n-Mayor}},\ }\href@noop {} {\bibfield  {journal} {\bibinfo
  {journal} {Physical Review E}\ }\textbf {\bibinfo {volume} {93}},\ \bibinfo
  {pages} {063308} (\bibinfo {year} {2016})}\BibitemShut {NoStop}%
\bibitem [{\citenamefont {Fishman}\ and\ \citenamefont
  {Aharony}(1979)}]{fishman1979random}%
  \BibitemOpen
  \bibfield  {author} {\bibinfo {author} {\bibfnamefont {S.}~\bibnamefont
  {Fishman}}\ and\ \bibinfo {author} {\bibfnamefont {A.}~\bibnamefont
  {Aharony}},\ }\href@noop {} {\bibfield  {journal} {\bibinfo  {journal}
  {Journal of Physics C: Solid State Physics}\ }\textbf {\bibinfo {volume}
  {12}},\ \bibinfo {pages} {L729} (\bibinfo {year} {1979})}\BibitemShut
  {NoStop}%
\bibitem [{\citenamefont {Hill}\ \emph {et~al.}(1993)\citenamefont {Hill},
  \citenamefont {Feng}, \citenamefont {Birgeneau},\ and\ \citenamefont
  {Thurston}}]{hill1993magnetic}%
  \BibitemOpen
  \bibfield  {author} {\bibinfo {author} {\bibfnamefont {J.~P.}\ \bibnamefont
  {Hill}}, \bibinfo {author} {\bibfnamefont {Q.}~\bibnamefont {Feng}}, \bibinfo
  {author} {\bibfnamefont {R.}~\bibnamefont {Birgeneau}},\ and\ \bibinfo
  {author} {\bibfnamefont {T.}~\bibnamefont {Thurston}},\ }\href@noop {}
  {\bibfield  {journal} {\bibinfo  {journal} {Zeitschrift f{\"u}r Physik B
  Condensed Matter}\ }\textbf {\bibinfo {volume} {92}},\ \bibinfo {pages} {285}
  (\bibinfo {year} {1993})}\BibitemShut {NoStop}%
\bibitem [{\citenamefont {Schechter}(2008)}]{schechter2008liho}%
  \BibitemOpen
  \bibfield  {author} {\bibinfo {author} {\bibfnamefont {M.}~\bibnamefont
  {Schechter}},\ }\href@noop {} {\bibfield  {journal} {\bibinfo  {journal}
  {Physical Review B—Condensed Matter and Materials Physics}\ }\textbf
  {\bibinfo {volume} {77}},\ \bibinfo {pages} {020401} (\bibinfo {year}
  {2008})}\BibitemShut {NoStop}%
\bibitem [{\citenamefont {Miga}\ \emph {et~al.}(2009)\citenamefont {Miga},
  \citenamefont {Kleemann}, \citenamefont {Dec},\ and\ \citenamefont
  {{\L}ukasiewicz}}]{miga2009three}%
  \BibitemOpen
  \bibfield  {author} {\bibinfo {author} {\bibfnamefont {S.}~\bibnamefont
  {Miga}}, \bibinfo {author} {\bibfnamefont {W.}~\bibnamefont {Kleemann}},
  \bibinfo {author} {\bibfnamefont {J.}~\bibnamefont {Dec}},\ and\ \bibinfo
  {author} {\bibfnamefont {T.}~\bibnamefont {{\L}ukasiewicz}},\ }\href@noop {}
  {\bibfield  {journal} {\bibinfo  {journal} {Physical Review B—Condensed
  Matter and Materials Physics}\ }\textbf {\bibinfo {volume} {80}},\ \bibinfo
  {pages} {220103} (\bibinfo {year} {2009})}\BibitemShut {NoStop}%
\bibitem [{\citenamefont {Westphal}\ \emph {et~al.}(1992)\citenamefont
  {Westphal}, \citenamefont {Kleemann},\ and\ \citenamefont
  {Glinchuk}}]{westphal1992diffuse}%
  \BibitemOpen
  \bibfield  {author} {\bibinfo {author} {\bibfnamefont {V.}~\bibnamefont
  {Westphal}}, \bibinfo {author} {\bibfnamefont {W.}~\bibnamefont {Kleemann}},\
  and\ \bibinfo {author} {\bibfnamefont {M.}~\bibnamefont {Glinchuk}},\
  }\href@noop {} {\bibfield  {journal} {\bibinfo  {journal} {Physical Review
  Letters}\ }\textbf {\bibinfo {volume} {68}},\ \bibinfo {pages} {847}
  (\bibinfo {year} {1992})}\BibitemShut {NoStop}%
\bibitem [{\citenamefont {Millis}\ \emph {et~al.}(2010)\citenamefont {Millis},
  \citenamefont {Kent}, \citenamefont {Sarachik},\ and\ \citenamefont
  {Yeshurun}}]{millis2010pure}%
  \BibitemOpen
  \bibfield  {author} {\bibinfo {author} {\bibfnamefont {A.}~\bibnamefont
  {Millis}}, \bibinfo {author} {\bibfnamefont {A.}~\bibnamefont {Kent}},
  \bibinfo {author} {\bibfnamefont {M.}~\bibnamefont {Sarachik}},\ and\
  \bibinfo {author} {\bibfnamefont {Y.}~\bibnamefont {Yeshurun}},\ }\href@noop
  {} {\bibfield  {journal} {\bibinfo  {journal} {Physical Review B—Condensed
  Matter and Materials Physics}\ }\textbf {\bibinfo {volume} {81}},\ \bibinfo
  {pages} {024423} (\bibinfo {year} {2010})}\BibitemShut {NoStop}%
\bibitem [{\citenamefont {Xu}\ \emph {et~al.}(2015)\citenamefont {Xu},
  \citenamefont {Silevitch}, \citenamefont {Dahmen},\ and\ \citenamefont
  {Rosenbaum}}]{xu2015barkhausen}%
  \BibitemOpen
  \bibfield  {author} {\bibinfo {author} {\bibfnamefont {J.}~\bibnamefont
  {Xu}}, \bibinfo {author} {\bibfnamefont {D.}~\bibnamefont {Silevitch}},
  \bibinfo {author} {\bibfnamefont {K.}~\bibnamefont {Dahmen}},\ and\ \bibinfo
  {author} {\bibfnamefont {T.}~\bibnamefont {Rosenbaum}},\ }\href@noop {}
  {\bibfield  {journal} {\bibinfo  {journal} {Physical Review B}\ }\textbf
  {\bibinfo {volume} {92}},\ \bibinfo {pages} {024424} (\bibinfo {year}
  {2015})}\BibitemShut {NoStop}%
\bibitem [{\citenamefont {Sreekala}\ and\ \citenamefont
  {Ananthakrishna}(2003)}]{sreekala2003acoustic}%
  \BibitemOpen
  \bibfield  {author} {\bibinfo {author} {\bibfnamefont {S.}~\bibnamefont
  {Sreekala}}\ and\ \bibinfo {author} {\bibfnamefont {G.}~\bibnamefont
  {Ananthakrishna}},\ }\href@noop {} {\bibfield  {journal} {\bibinfo  {journal}
  {Physical review letters}\ }\textbf {\bibinfo {volume} {90}},\ \bibinfo
  {pages} {135501} (\bibinfo {year} {2003})}\BibitemShut {NoStop}%
\bibitem [{\citenamefont {Sreekala}\ \emph {et~al.}(2004)\citenamefont
  {Sreekala}, \citenamefont {Ahluwalia},\ and\ \citenamefont
  {Ananthakrishna}}]{sreekala2004precursors}%
  \BibitemOpen
  \bibfield  {author} {\bibinfo {author} {\bibfnamefont {S.}~\bibnamefont
  {Sreekala}}, \bibinfo {author} {\bibfnamefont {R.}~\bibnamefont
  {Ahluwalia}},\ and\ \bibinfo {author} {\bibfnamefont {G.}~\bibnamefont
  {Ananthakrishna}},\ }\href@noop {} {\bibfield  {journal} {\bibinfo  {journal}
  {Physical Review B—Condensed Matter and Materials Physics}\ }\textbf
  {\bibinfo {volume} {70}},\ \bibinfo {pages} {224105} (\bibinfo {year}
  {2004})}\BibitemShut {NoStop}%
\bibitem [{\citenamefont {Vives}\ \emph {et~al.}(1994)\citenamefont {Vives},
  \citenamefont {Ort{\'\i}n}, \citenamefont {Ma{\~n}osa}, \citenamefont
  {R{\`a}fols}, \citenamefont {P{\'e}rez-Magran{\'e}},\ and\ \citenamefont
  {Planes}}]{vives1994distributions}%
  \BibitemOpen
  \bibfield  {author} {\bibinfo {author} {\bibfnamefont {E.}~\bibnamefont
  {Vives}}, \bibinfo {author} {\bibfnamefont {J.}~\bibnamefont {Ort{\'\i}n}},
  \bibinfo {author} {\bibfnamefont {L.}~\bibnamefont {Ma{\~n}osa}}, \bibinfo
  {author} {\bibfnamefont {I.}~\bibnamefont {R{\`a}fols}}, \bibinfo {author}
  {\bibfnamefont {R.}~\bibnamefont {P{\'e}rez-Magran{\'e}}},\ and\ \bibinfo
  {author} {\bibfnamefont {A.}~\bibnamefont {Planes}},\ }\href@noop {}
  {\bibfield  {journal} {\bibinfo  {journal} {Physical review letters}\
  }\textbf {\bibinfo {volume} {72}},\ \bibinfo {pages} {1694} (\bibinfo {year}
  {1994})}\BibitemShut {NoStop}%
\bibitem [{\citenamefont {Vives}\ \emph {et~al.}(1995)\citenamefont {Vives},
  \citenamefont {R{\`a}fols}, \citenamefont {Ma{\~n}osa}, \citenamefont
  {Ort{\'\i}n},\ and\ \citenamefont {Planes}}]{vives1995statistics}%
  \BibitemOpen
  \bibfield  {author} {\bibinfo {author} {\bibfnamefont {E.}~\bibnamefont
  {Vives}}, \bibinfo {author} {\bibfnamefont {I.}~\bibnamefont {R{\`a}fols}},
  \bibinfo {author} {\bibfnamefont {L.}~\bibnamefont {Ma{\~n}osa}}, \bibinfo
  {author} {\bibfnamefont {J.}~\bibnamefont {Ort{\'\i}n}},\ and\ \bibinfo
  {author} {\bibfnamefont {A.}~\bibnamefont {Planes}},\ }\href@noop {}
  {\bibfield  {journal} {\bibinfo  {journal} {Physical Review B}\ }\textbf
  {\bibinfo {volume} {52}},\ \bibinfo {pages} {12644} (\bibinfo {year}
  {1995})}\BibitemShut {NoStop}%
\bibitem [{\citenamefont {Dahmen}\ \emph {et~al.}(1994)\citenamefont {Dahmen},
  \citenamefont {Kartha}, \citenamefont {Krumhansl}, \citenamefont {Roberts},
  \citenamefont {Sethna},\ and\ \citenamefont {Shore}}]{dahmen1994disorder}%
  \BibitemOpen
  \bibfield  {author} {\bibinfo {author} {\bibfnamefont {K.}~\bibnamefont
  {Dahmen}}, \bibinfo {author} {\bibfnamefont {S.}~\bibnamefont {Kartha}},
  \bibinfo {author} {\bibfnamefont {J.~A.}\ \bibnamefont {Krumhansl}}, \bibinfo
  {author} {\bibfnamefont {B.~W.}\ \bibnamefont {Roberts}}, \bibinfo {author}
  {\bibfnamefont {J.~P.}\ \bibnamefont {Sethna}},\ and\ \bibinfo {author}
  {\bibfnamefont {J.~D.}\ \bibnamefont {Shore}},\ }\href@noop {} {\bibfield
  {journal} {\bibinfo  {journal} {Journal of Applied Physics}\ }\textbf
  {\bibinfo {volume} {75}},\ \bibinfo {pages} {5946} (\bibinfo {year}
  {1994})}\BibitemShut {NoStop}%
\end{thebibliography}%
\bibliographystyle{apsrev4-2}

\end{document}